\documentclass[letterpaper,journal,romanappendices]{IEEEtran}
\usepackage{xcolor}
\usepackage{amsmath,amssymb,amsfonts,amsthm}
\usepackage{algorithmic}
\usepackage{algorithm}
\usepackage{array}
\usepackage{textcomp}
\usepackage{stfloats}
\usepackage{url}
\usepackage{verbatim}
\usepackage{graphicx}
\usepackage[caption=false,font=footnotesize,
labelfont=sf,textfont=sf]{subfig}
\usepackage{cite}
\usepackage{xcolor}
\usepackage{braket}
\usepackage[hidelinks]{hyperref}
\usepackage[T1]{fontenc}
\usepackage{cleveref}
\usepackage{gensymb}
\usepackage{orcidlink}
\usepackage{braket}
\usepackage{subfig}
\def\BibTeX{{\rm B\kern-.05em{\sc i\kern-.025em b}\kern-.08em
    T\kern-.1667em\lower.7ex\hbox{E}\kern-.125emX}}

\newcommand{\dif}{\mathop{}\!\mathrm{d}}
\newcommand{\im}{j}
\newcommand{\N}{10^3}

\allowdisplaybreaks

\renewcommand{\thesubsubsection}{\arabic{subsubsection}}

\newcommand{\revFinal}[1]{#1}
\begin{document}

\title{Experimental Validation of Provably Covert Communication Using Software-Defined Radio
\thanks{This work was supported, in part, by the National Science Foundation under Grants No.~CCF-2006679 and CNS-2107265.}
}

\author{Rohan Bali\orcidlink{0009-0007-7992-8803}, Trevor E. Bailey\orcidlink{0009-0006-7516-5100}, Michael S.~Bullock\orcidlink{0000-0002-3528-7473}, and Boulat A.~Bash\orcidlink{0000-0002-1205-3906}
\thanks{The authors are with the Department of Electrical and Computer Engineering at the University of Arizona, Tucson, AZ, USA (email: \{rbali, trevorbailey, bullockm, boulat\}@arizona.edu)}}

\maketitle

\begin{abstract}
The fundamental information-theoretic limits of covert, or low probability of detection/intercept (LPD/LPI), communication have been extensively studied for over a decade, resulting in the \emph{square root law} (SRL): only $L\sqrt{n}$ covert bits can be reliably transmitted over time-bandwidth product $n$, for constant $L>0$.
Transmitting more either results in detection or decoding errors.
The SRL imposes significant constraints on the hardware realization of mathematically guaranteed covert communication. These preclude the use of standard link maintenance operations that are taken for granted in non-covert communication.
Thus, experimental validation of covert communication is underexplored: to date, only two experimental studies of SRL-based covert communication are available, both focusing on optical channels. 
Here, we demonstrate provably secure covert radio-frequency (RF) communication using software-defined radios (SDRs).
We combine system design, theory, and experiments by 1) developing a sparse-signaling pulse shape to enable covert transmission of data in an environment with potential mobility, 2) proving the covertness of the resulting system, and 3) validating the theoretical predictions by implementing it on SDRs.
We uncover and address unique challenges specific to using SDR architecture for covert communications.
This opens practical avenues for implementing covert communication systems and raises further research questions.
\end{abstract}
\section{Introduction}
\label{sec:introduction}
Covert, or low probability of detection/intercept (LPD/LPI) communication, allows transmitting messages without alerting an adversary \cite{bash12sqrtlawisit, bash13squarerootjsacnonote, bash15covertcommmag, chen23covcommssurvey}. This mode of information security is highly desirable  and contrasts with the traditional cryptographic \cite{menezes96HAC} and information-theoretic secrecy \cite{bloch11pls} methods that prevent adversaries' access to the transmission's content, but not its detection.
Careful waveform design and spread spectrum techniques are often employed in practice to reduce adversaries' signal-to-noise ratio (SNR) below the noise floor \cite[Pt.~1, Ch.~5]{simon94ssh}.
However, mathematically guaranteeing covertness requires following the \emph{square root law} (SRL): only $B(n)=L\sqrt{n}$ covert bits can be reliably transmitted over $n$ channel uses \cite{bash12sqrtlawisit, bash13squarerootjsacnonote, bash15covertcommmag, chen23covcommssurvey}.
The channel-dependent constant $L>0$ is called \emph{covert capacity} and $n=TW$ is typically the time-bandwidth product for the transmission.
Notably, the Shannon capacity \cite{cover02IT} of this channel is zero, since $\lim_{n\to\infty}\frac{B(n)}{n}=0$.
This is because the adversary's task in covert communication involves learning just one bit of information (whether the transmitter is on or not) versus $\mathcal{O}(n)$ bits of transmitted data in traditional secure communication.
Nevertheless, a significant number of such mathematically guaranteed covert bits can still be transmitted within a finite time.

The discovery of the SRL in \cite{bash12sqrtlawisit, bash13squarerootjsacnonote} resulted in an explosion of research by the communication and information theory communities, as overviewed in a tutorial \cite{bash15covertcommmag} and a detailed survey \cite{chen23covcommssurvey}.
This included characterization of capacity $L$ for additive white Gaussian noise (AWGN) and discrete memoryless channels (DMCs) \cite{bloch15covert, wang15covert, tahmasbi19covertdmc2ndorder, xinchun24secondorderawgncovert}, covert networks \cite{arumugam16broadcast,  hu17covertnetworks, hu18greedyrelay, arumugam18mac, tan18covertbc, azadeh18covertmultihop, soltani18netlpd, zheng19multiantennacovertcommswirelessnets, liu20iotthz, cho20covertscaling-journal, hieu23ratesplittingcovertmac, kong24covertroutinghetnets, zhao25wardencollusioncovertnets}, quantum aspects of covert communication \cite{bash13quantumlpdisit, bash15covertbosoniccomm, azadeh16quantumcovert-isitarxiv, bullock20discretemod, gagatsos20codingcovcomm, tahmasbi19bosoniccovertqkd, tahmasbi20bosoniccovertqkd-jsait,
anderson21bosonicbroadast, wang22isitcoverttd, anderson2024covert-qce, bullock2025fundamentallimitscovertcommunication, zlotnick25eacovertcomm}, covert communication with unmanned aerial vehicles (UAVs) \cite{jiang21covcommUAVnets, chen21uavjammer, wang23covcommUAV-IRS, tian24uavcovcommasymetricinfo, xu25multiuserUAV, wu25covertUAVsurvey}, and many other directions outlined in recent surveys \cite{chen23covcommssurvey, wu25covertUAVsurvey, li25intelligentcovcomm} and references therein.

An adversary lacking knowledge can improve covert communication.
For example, an adversary's uncertainty of transmission time/frequency yields a multiplicative enhancement to covert capacity $L$ \cite{bash14timingisit, bash16timingtwc, arumugam16async}.
Uncertainty in adversarial channel noise power levels, from, e.g., jammer assistance, can, in theory, lead to a linear law: $\mathcal{O}(n)$ covert bits reliably transmissible over $n$ channel uses \cite{sobers15jammer-asilomar, sobers17jammer, zheng19multiantenna, chen21uavjammer, zheng21distributedjammingcovcomm}.
However, one must approach the assumptions underlying these results with care, e.g., if the underlying channel is continuous-time, transmissions that break the SRL are detectable \cite{gillani23covertcommsconttime}.

On the contrary, the SRL arguably governs the worst case by assuming the adversary knows the transmitter-adversary channel's characteristics, exact transmission time/frequency, and transceiver design. As detailed in Section \ref{sec:prerequisite}, the adversary's only limitations are: 1)  inability to control the random channel noise; and 2) lack of access to the secret shared between transmitter and receiver prior to the transmission.
Therefore, the SRL, while conservative, provides a high level of security against unforeseen adversarial technological surprises.
Unfortunately, the strict output power constraint of SRL-based covert communication precludes the use of standard handshakes, carrier synchronization, channel state estimation, and other link-maintenance operations that are taken for granted in non-covert communication. 
Hence, notwithstanding the progress on fundamental theory, experimental covert communication remains underexplored, with few published works focusing on optical channels \cite{bash15covertbosoniccomm, liu24metrofibercovcomm, Djordjevic25covertSLMatmospheric, Djordjevic25covertSLM}.

This paper, to our knowledge, provides the first experimental validation of the SRL on radio frequency (RF) channels.
We implement and evaluate a covert communication protocol using software-defined radio (SDR).
Section \ref{sec:Experiment_Setup} describes our implementation, which uses USRP X310 SDR units on the COSMOS (formerly ORBIT) testbed \cite{raychaudhuri05orbit, orbit_grid}. This enables controlled and reproducible experiments.
We employ quadrature phase-shift keying (QPSK) with a Gaussian pulse shaping filter.
This controls temporal and spectral symbol leakage while mitigating timing jitter.
We implement an experimental framework based on a synthetic Gaussian noise source within a wired, shielded network that provides environmental control and supports reproducibility.

Maintaining covert links necessitates using precise timing sources, such as atomic clocks.
Furthermore, the finite resolution of modern digital transceivers' (such as SDRs) digital-to-analog and analog-to-digital converters (DACs and ADCs) imposes a fundamental \emph{lower} limit on the amplitude of the transmitted symbol.
Thus, to meet the SRL, we must use a ``sparse coding'' strategy, selecting a random subset of available channel uses that is secretly shared with the receiver in advance.
This sparsity, in turn, renders the standard phase synchronization approaches that rely on periodic pilot signal transmission unusable.
Therefore, we include a pilot signal with \emph{every} transmitted symbol, which is done rarely if at all in non-covert communications.
This, along with precise time synchronization, enables our SDR implementation to transmit a significant number of bits with a guarantee of covertness.

While we consider the worst-case adversary with complete access to the covert communication system's clock, our experiment motivates a systematic study of the impact of adversary hardware limitations, such as sampling timing jitter, as well as receiver bandwidth constraints.
Furthermore, employing more efficient modulation and coding schemes (such as \cite{wang21covcodes}), as well as minimizing the amount of pre-shared secret (see \cite{bloch15covert}), are avenues for future exploration.
Finally, other intermittent communication systems may benefit from our approaches.

This paper is organized as follows: Section~\ref{sec:prerequisite} develops the theoretical foundation for radio-frequency covert communication, beginning with the discrete-time AWGN channel model and deriving a sparse-coded QPSK transmission scheme that satisfies the SRL. 
Section~\ref{sec:implementation} details our SDR implementation on the COSMOS testbed and describes the experimental procedure, while Section \ref{sec:results} presents measurement results that validate the theory.
Section~\ref{sec:discussion} interprets these findings, highlighting practical design challenges and their implications for real-world systems, and outlines open problems and promising directions for extending this work.
Finally, Appendices provide supporting derivations for the covertness criterion, phase-offset estimation, and SNR estimation used in the experiments.

\section{Analysis of Practical Covert Communications} \label{sec:prerequisite}

\subsection{Channel Model}
\label{subsec:channel_model}
\begin{figure}
    \centering
    \includegraphics[width=1\linewidth]{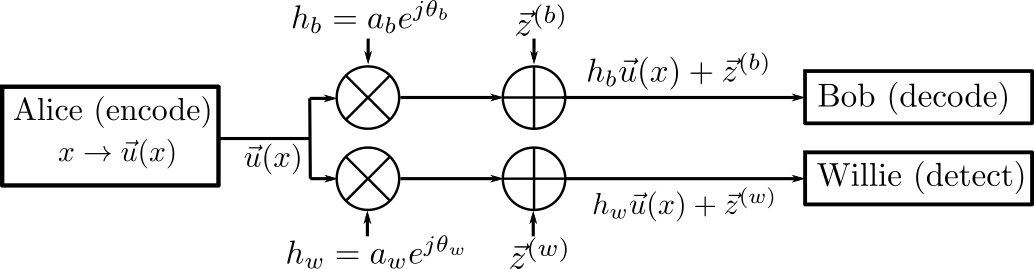}
    \caption[Discrete-time AWGN block broadcast channel.]{Discrete-time AWGN block broadcast channel. Alice modulates $x\in\mathcal{X}$ into a complex-valued pulse shape $\vec{u}(x)\in\mathbb{C}^{n_s}$ and transmits over $n_s$ uses of a channel corrupted by independent AWGN at legitimate receiver Bob and adversary (warden) Willie. Bob receives $h_{b}\vec{u}(x) + \vec{z}^{(b)}$ while Willie observes $h_{w}\vec{u}(x) + \vec{z}^{(w)}$, where the channel gain  $h_{r}=a_re^{\im\theta_r}$ from Alice to the receiver $r\in\{b,w\}$ contains attenuation $a_r>0$ and a random phase $\theta_r$. We model noise $\vec{z}^{(r)}$ by circularly-symmetric complex Gaussian random vectors.}
    \label{fig:theoretical_channel}
\end{figure}

As our SDRs are digital and time-synchronized (see Section \ref{sec:Experiment_Setup} for details), the discrete-time AWGN block channel model in Fig.~\ref{fig:theoretical_channel} describes their input-output relationship in our physical system.
A channel use corresponds to a discrete sample of Alice's input, corrupted by loss and AWGN, as observed by the legitimate receiver Bob and adversary (warden) Willie.
Alice uses a discrete-time $n_s$-length complex-valued pulse shape $\vec{u}(x)\in\mathbb{C}^{n_s}$ to modulate data in $x\in\mathcal{X}$, where the modulation constellation $\mathcal{X}$ is defined in Section \ref{sec:system_design}.
She transmits this pulse over $n_s$ channel uses to  Bob and Willie.
Bob receives $h_{b}\vec{u}(x) + \vec{z}^{(b)}$ while Willie observes $h_{w}\vec{u}(x) + \vec{z}^{(w)}$.
The respective complex-valued channel gains $h_{b}=a_b e^{\im\theta_b}$ and $h_{w}=a_w e^{\im\theta_w}$ contain attenuation $a_b,a_w>0$ and phase $\theta_b,\theta_w\in[0,2\pi]$ components.
We assume channel coherence time longer than the pulse duration; thus, the channel gains $h_b$ and $h_w$ are constant for at least $n_s$ AWGN channel uses.
For each pulse, attenuation is arbitrary, while phase is assumed uniformly random in $[0,2\pi]$.
Each symbol in the transmitted pulse $\vec{u}(x)$ is corrupted independently by AWGN: $\vec{z}^{(b)}$ and $\vec{z}^{(w)}$ are instances of independent and identically distributed (i.i.d.) circularly-symmetric complex Gaussian random vectors $\vec{Z}^{(b)}\sim\mathcal{CN}\left(\vec{0},2\sigma_b^2\mathbf{I}_{n_s}\right)$ and $\vec{Z}^{(w)}\sim\mathcal{CN}\left(\vec{0},2\sigma_w^2\mathbf{I}_{n_s}\right)$, where $\mathbf{I}_{n_s}$ is an $n_s\times n_s$ identity matrix.

The pulse-to-pulse variation in channel gain models the dynamic operating environment caused by the mobility of Alice, Bob, and Willie, as well as the surrounding reflecting objects.
Assigning a suitable probability distribution (e.g., Rayleigh, Rice, or Nakagami) to attenuation coefficients $a_b$ and $a_w$ (along with uniform phase distribution) yields a corresponding block fading model (see \cite[Ch.~2]{tse05wirelesscomm}).
Phases $\theta_b$ and $\theta_w$ are random for each pulse slot, which captures not only environment dynamics but also the drift of independent oscillators. 
However, for simplicity, we assume that attenuation is constant across all channels.
This models a scenario where, say, worst-case path-loss figures are employed. We defer the analysis of fully-dynamic channels to future work.

\subsection{Practical Covert System Design}
\label{sec:system_design}

\begin{figure*}[!t]
\centering
\includegraphics[width=\textwidth]{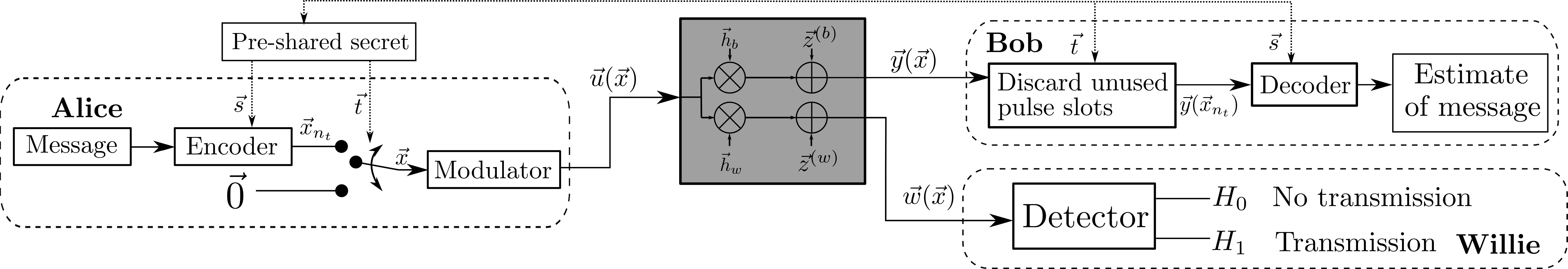}
\caption{
System model for covert communication. Alice either encodes, modulates, and transmits a message in $n_p$ pulses over $n = n_pn_s$ uses of the AWGN channel, or she remains silent. Here, $\vec{x}\in\left(\mathcal{X}\cup\{0\}\right)^{n_p}$ is the vector of QPSK and zero symbols, with the relative location of each selected using $\vec{t}$. After modulation, $\vec{u}(\vec{x})\in \mathbb{C}^n$ is the input pulse-shaped encoded signal from Alice. AWGN channel is characterized by additive noise $\vec{z}^{(w)}\in\mathbb{C}^{n}$ and channel gain $\vec{h}_b,\vec{h}_w \in \mathbb{C}^{n_p}$, as depicted in Fig.~\ref{fig:theoretical_channel}.  Here, each element of $\vec{h}_b,\vec{h}_w$ applies uniformly to one $n_s$ sample pulse slot. Then, $\vec{y}(\vec{x})\in\mathbb{C}^n$ (resp.~$\vec{w}(\vec{x})\in\mathbb{C}^n$) is the output at Bob (resp.~Willie). Willie's goal is to determine whether Alice is transmitting.  Bob and Alice use a pre-shared secret $(\vec{s},\vec{t})$ to prevent this while communicating reliably.}
\label{fig:system_model}
\end{figure*}

The \emph{square root law} (SRL) governs covert communication over the AWGN channel in Fig.~\ref{fig:theoretical_channel}: one can transmit $B(n)=L\sqrt{n}$ covert bits reliably in $n$ uses of such channel \cite{bash12sqrtlawisit, bash13squarerootjsacnonote}, where $L>0$ is the covert capacity (derived in \cite{bloch15covert, wang15covert}) and $n=TW$ is the transmission time-bandwidth product.
These results and much of the follow-up employ a simplified scenario with $n_s=1$, $a_b=a_w=1$, and $\theta_b=\theta_w=0$.
However, practical radio systems require pulse shaping to efficiently use available bandwidth and mitigate inter-symbol interference and timing jitter.
Furthermore, analysis of the fundamental limits often assumes that Alice can generate inputs at arbitrarily low power.
In practice, DACs and ADCs have finite resolution, which imposes a lower limit on the output signal power.
We address this next, and then discuss modulation and pulse shape.
Our overall system model is depicted in Fig.~\ref{fig:system_model}. 

\subsubsection{Sparse Coding}
\label{subsec:sparse_coding}
Divide $n$ available channel uses into $n_p$ pulses, with shape given by the $n_s$-length pulse-shape vector $\vec{u}(x)$ discussed next, $x\in\mathcal{X}$.
Then, $n=n_pn_s$, with $n_p\in\mathcal{O}(n)$.
The minimum output power limit implies the norm $\|\vec{u}(x)\|=c$ for all $x\in\mathcal{X}$. We use \emph{sparse coding} to ensure covertness: before transmission, Alice and Bob randomly generate and secretly share an $n_p$-length sequence $\vec{t}$ of i.i.d.~samples from the Bernoulli distribution $p(t_i)=\{
1-\alpha_n \text{ if } t_i = 0; \alpha_n \text{ if } t_i = 1\}$, where the probability of transmitting in a given slot $\alpha_n \in \mathcal{O}\left(1/\sqrt{n}\right)$ follows the SRL \cite{bash12sqrtlawisit, bash13squarerootjsacnonote}. 
The number $n_t=\sum_{i=1}^{n_p} t_i$ of selected pulse slots is a random variable with mean $\alpha_nn_p$. It is also the length of the transmitted message $\vec{x}_{n_t}$, in symbols.
Alice is quiet in the slots that are not selected, that is, when $t_i=0$, she inputs $\vec{0}$.
We assume that Bob and Alice are synchronized by a common clock.
Denoting by $x=0$ this ``innocent'' symbol, the transmitted symbol vector $\vec{x}\in\left(\mathcal{X}\cup\{0\}\right)^{n_p}$ contains $n_t$ symbols from $\mathcal{X}$ and $n_p-n_t$ innocent symbols.
This yields Alice's sparse-coded covert input $\vec{u}(\vec{x})\in\mathbb{C}^n$, containing mostly innocent symbols.
We derive $\alpha_n$ in Section \ref{subsec:covertness}.

\subsubsection{Modulation and Pulse Shape}
\label{subsec:covert_pulse_design}

The channel model in Section \ref{subsec:channel_model} includes random pulse-to-pulse phase variation.
On-off keying (OOK) modulation is resilient to this as it uses energy pulses. However, OOK is power-inefficient; matching the per-symbol bit-error rate (BER) of binary and quadrature phase-shift keying (BPSK and QPSK) schemes requires transmitting twice the power, which, in turn, doubles Willie's received SNR and harms covertness. We discuss the relationship between SNR and covert signal construction further in Section \ref{subsec:covertness}.
We choose QPSK modulation since it doubles the number of degrees of freedom relative to BPSK, further increasing the achievable covert communication rate.

\begin{figure}[t]
  \label{fig:pulse_shape}
  \centering
  \includegraphics[width=\linewidth]{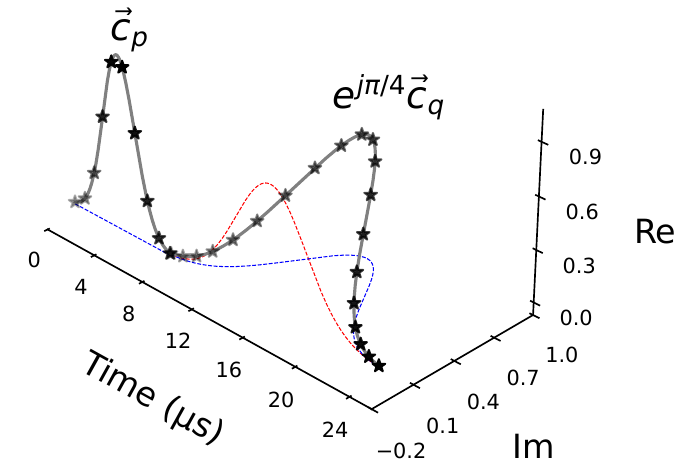}
  \caption{Pulse shape $\vec{u}(1)$. A strictly positive pilot segment $\vec{c_p}$ precedes the data segment $\vec{c_q}$, whose in-phase and quadrature components form the QPSK symbol $e^{\frac{\im\pi}{4}}$ encoding $x=1$. The projections of the complex waveform onto the real and imaginary axes are overlaid with dotted red and blue lines, respectively, and the discrete time instants at which the Gaussian envelope is sampled for transmission are marked by $\star$. 
  \label{fig:one_pulse_slot}}
\end{figure}

Phase variations cause errors in phase-keyed systems and are typically mitigated using periodic pilot signals; however, the SRL does not permit periodic transmissions. Furthermore, the sparse coding scheme results in the time between pulses scaling as $\mathcal{O}(\sqrt{n})$ on average, rendering the phases uncorrelated for sufficiently large $n$ when the channel coherence time is finite. Thus, we include a fixed $n_s^{(p)}$-length strictly-positive pilot segment $\vec{c}_p$ in \emph{every} transmitted pulse. For sufficiently long channel coherence time, the channel induces approximately the same phase rotation on both the pilot and data segments, preserving their relative phase. The pilot segment therefore allows estimation of the phase to correct the $n_s^{(q)}$-length QPSK-modulated data segment $\vec{c}_q(x)=e^{\frac{\im \pi (2x-1)}{4}}\vec{c}_q$, $x\in\mathcal{X}$, where $\mathcal{X}=\{1,\ldots,4\}$ and each $x\in\mathcal{X}$ is represented by two bits. Alice additionally applies a random phase $\theta \sim \mathcal{U}[0,2\pi)$ to both pilot and QPSK pulse in each transmitted pulse. This improves covertness by ensuring that the relative phase between transmissions is randomized without any additional cost to decoding reliability, since Bob needs to estimate the phase anyway. Hence, the transmitted pulse is given by $\vec{u}(x)=\left(e^{j\theta}\vec{c}_p,e^{j\theta}\vec{c}_q(x)\right)$ as depicted in Fig.~\ref{fig:pulse_shape}, with length $n_s=n_s^{(p)}+n_s^{(q)}$ and norm $\|\vec{u}(x)\|\triangleq c$ for all $x\in\mathcal{X}$. Since the pilot requires additional power, this scheme increases Willie's received SNR. Thus, there is a trade-off between phase estimation and the covertness of our signal that we optimize in Section \ref{subsec:hyperparam_opt}.

Sparse coding has negligible inter-symbol interference.
Instead, covertness demands energy confinement in time and frequency domains.
Thus, we shape both  $\vec{c}_p$ and $\vec{c}_q$ by a discrete-time Gaussian envelope: 
\begin{equation}
  \vec{g}_k[m]
  = \exp\left({-\frac{\left(m-n_s^{(k)}/2\right)^{2}}{2\sigma_k^{2}}}\right),
  \quad
  \left.
  \begin{aligned}
    m &= 0,\dots,n_s^{(k)}-1,\\
    k &\in \{p,q\}
  \end{aligned}
  \right.
  \label{eq:gaussian_env}
\end{equation} 
and define
\begin{align}
    \vec{c}_k \triangleq c_k\frac{ \vec{g}_k }{\|\vec{g}_k\|},  \quad
  \left.
  \begin{aligned}
    k &\in \{p,q\}
  \end{aligned}
  \right.
\end{align}
where $\sigma_p$, $\sigma_q$, $c_p$, and $c_q$ are the corresponding pulse width and magnitude parameters. We note that more complicated pulse shapes with lower bandwidth under the same power and time constraints exist, e.g., discrete prolate spheroidal (Slepian) sequences \cite[Sec.~3.1]{slepian78discreteprolate}
but we defer using them to future work.
In Section \ref{subsec:hyperparam_opt}, we optimize $n_s^{(p)}$, $n_s^{(q)}$, $c_p$, and $c_q$. Thus, we choose the largest $\sigma_k$ satisfying
\begin{equation}
  \frac{\sum_{m=0}^{n_s^{(k)}-1} \vec{c}_{k}[m]}{\sum_{m=-\infty}^{\infty} \vec{c}_{k}[m]} \;\ge\; 0.999,
  \qquad k\in\{p,q\},
  \label{eq:sigma_criterion}
\end{equation}
which ensures that at least $99.9\%$ of each pulse's energy is confined to its designated slot. 

\subsection{Reliability Analysis}
\label{subsec:sysmodel}

Bob has $\vec{t}$ and, thus, can discard the slots that are not used for transmission (that is, those slots corresponding to $x=0$ in $\vec{x}$ when Alice ``transmits'' $\vec{0}$).
Thus, Bob only examines $n_t$ pulses in $\vec{y}(\vec{x}_{n_t})$.
When Alice transmits a symbol $x \in \mathcal{X}$ from $\vec{x}_{n_t}$ in each of the $n_t$ selected pulse slots, Bob receives the $n_s$-sample vector $\vec{y}=\left(\vec{y}_p,\vec{y}_q\right)$ containing two segments:
\begin{align}
\vec{y}_p&=a_be^{\im\theta_b} \vec{c}_p + \vec{z}^{(b)}_p\\
\vec{y}_q(x)&=a_be^{\frac{\im\pi (2x-1)}{4}+\theta_b} \vec{c}_q + \vec{z}^{(b)}_q
\end{align}
where $a_b$ is a constant loss, $\theta_b$ is unknown phase offset uniformly-distributed from $[0,2\pi]$, and $\vec{z}^{(b)}=(\vec{z}^{(b)}_p,\vec{z}^{(b)}_q)$ is complex-valued circularly-symmetric AWGN, per Section \ref{subsec:channel_model}.
Bob uses $\vec{y}_p$ to obtain the estimate $\hat{\theta}_b$ of  $\theta_b$ 
(see Appendix~\ref{ap:phase_est}), and correct the phase of the data segment by applying $\vec y_q(x)e^{-\im\hat{\theta}_b}$. Bob decodes the QPSK symbol by first projecting the phase-aligned data segment onto the pulse shape,
\begin{align}
r_{\mathrm I}
  &= \left\langle\Re\{\vec{y}_q(x)\,e^{-j\hat{\theta}_b}\},
              \vec{c}_q\right\rangle,
&
r_{\mathrm Q}
  &= \left\langle\Im\{\vec{y}_q(x)\,e^{-j\hat{\theta}_b}\},
              \vec{c}_q\right\rangle.
\end{align}
This yields two BPSK symbols. Hard decisions $\hat{\tilde b}_1=\operatorname{sgn}(r_{\mathrm I})$ and $\hat{\tilde b}_0=\operatorname{sgn}(r_{\mathrm Q})$, followed by $\hat{b}_k=(1-\hat{\tilde{b}}_k)/2$, $k\in\{0,1\}$, extract the two bits.
Since AWGN is symmetric, this hard-decision scheme induces two binary symmetric channels (BSCs) with a probability of error: 
\begin{align}
     p_{e,\mathrm{bsc}}^{(b)}
  \triangleq
  \Pr(\hat{b}_k=0\mid b_k=1)
  =\Pr(\hat{b}_k=1\mid b_k=0).\label{eq:peb}
\end{align}
For sufficiently large $n_p$, Alice and Bob can use an error correction code (ECC) \cite{richardson2008moderncoding} on the approximately $\alpha_n n_p$-long subset of pulse slots $\{v:t_v=1\}$. 
This allows reliable transmission of
\begin{align}
    B_{\rm bsc}(n)&= 2n_tC_{\rm bsc} \approx 2\alpha_n n_p C_{\rm bsc}\label{eq:tp} 
\end{align}
bits in $n$ channel uses on average, with $C_{\rm bsc}\triangleq 1-h_2\left(p_{e, \rm{bsc}}^{(b)}\right)$, where $h_2(p)\triangleq-p\log_2(p)-(1-p)\log_2(1-p)$ is the binary entropy function and the approximation is due to $n_t$ being a random variable.

The ECC structure, which can aid Willie in detection, is eliminated by applying a $2n_t$-bit one-time pad $\vec{s}$ to the binary encoder output.
Each bit in $\vec{s}$ is selected by Alice and Bob equiprobably at random, resulting in output distribution $\Pr(x=-1)=\Pr(x=1)=\frac{1}{2}$. Pre-shared secret includes $\vec{t}$ and $\vec{s}$.

We require pre-sharing of $\mathcal{O}(\sqrt{n}\log n)$ secret key bits, since $\vec{t}$ and $\vec{s}$ take $n_t\log_2 n_p$ and $2n_t$ bits to represent, respectively. 
It is possible to reduce the required number of secret bits to $\mathcal{O}(\sqrt{n})$ at a significant complexity expense \cite{bloch15covert}.
As computation uses more energy than storage, in the context of future portable covert communication systems, the logarithmic cost is worth the simplicity of our scheme.

\subsection{Hypothesis Testing and Covertness} \label{subsec:covertness}
Willie observes the sequence $\vec{w}\in\mathbb{C}^n$ and must decide whether Alice is transmitting.
Unlike Bob, he does not have access to $\vec{t}$ and $\vec{s}$, but he knows the exact transmission start time and pulse boundaries.
That is, Willie can access Alice and Bob's shared time, which we model in the experiment by using a common clock for all parties.
Additionally, we assume that he knows the state of his channel from Alice, and other details of Alice and Bob's transceiver design (including $\vec{c}_p$, $\vec{c}_q$, and $\alpha_n$). 
The stringency of these assumptions ensures that the system is secure against the most capable adversary.
We defer studying the impact of relaxing these to future work.

Willie performs a test between binary hypotheses $H_0$ (silence) and $H_1$ (transmission). 
Let $P_{\text{FA}}=P(\text{choose~}H_1|H_0)$ and $P_{\text{MD}}=P(\text{choose~}H_0|H_1)$ denote the probabilities of false alarm and missed detection, respectively.
Assuming non-informative priors\footnote{Accounting for arbitrary priors is discussed in \cite{sobers17jammer}.} $\Pr(H_0)=\Pr(H_1)=\frac{1}{2}$, Willie's probability of error is:
\begin{align}
p_e^{(w)}&\triangleq \frac{1}{2}\left(P_{\text{FA}}+P_{\text{MD}}\right).\label{eq:pew}
\end{align}
Let distributions $P_0^{n}$ and $P_1^{n}$ and associated density functions $p_0^n(\vec{w})$ and $p_1^n(\vec{w})$ describe the statistics of Willie's output $\vec{w}$ when Alice is silent ($H_0$) and transmitting ($H_1$).
An optimal detection scheme yields \cite[Th.~15.1.1]{lehmann05stathyp4ed}:
\begin{align}
    p_e^{(w)}=\frac{1}{2}-\frac{1}{2}\mathcal{V}_T(P_0^{n},P_1^n),\label{eq:tvbound}
\end{align}
where $\mathcal{V}_T(P_0^{n},P_1^n)\triangleq\frac{1}{2}\int_{\mathbb{V}^n}\dif^2\vec{w}|p_0^n(\vec{w})-p_1^n(\vec{w})|$ is the total variation distance between $P_0^{n}$ and $P_1^{n}$.
We say that the transmission is $\delta$-covert if $p_e^{(w)}\geq\frac{1}{2}-\delta$.
Total variation distance is mathematically unwieldy, 
hence, in covert communication literature \cite{bash12sqrtlawisit, bash13squarerootjsacnonote,bloch15covert,wang15covert} we often employ Pinsker's inequality \cite[Lemma 11.6.1]{cover02IT} to lower bound 
\begin{align}
p_e^{(w)}
    &\geq \frac{1}{2}-\frac{1}{2\sqrt{2}}\sqrt{D(P_0^{ n}\|P_1^n)},\label{eq:pinskerbound}
\end{align}
where $D(P_0^{ n}\|P_1^n)\triangleq\int_{\mathbb{C}^n}\dif^2\vec{w}p_0^n(\vec{w})\log_2\frac{p_0^n(\vec{w})}{p_1^n(\vec{w})}$ is the relative entropy of $P_0^{n}$ and $P_1^{n}$. 
Thus, instead of \eqref{eq:tvbound}, one can use \eqref{eq:pinskerbound}, and assert that any scheme is $\delta$-covert if $D(P_0^{ n}\|P_1^n)\leq\delta_{\rm RE}=8\delta^2$. Ultimately, 
satisfying this relative entropy constraint for a given $\delta$ requires limiting Alice's average output power. While relative entropy between our distributions can be readily bounded, Pinsker's 
inequality is loose for many distribution pairs. 
Thus, we seek a tighter bound than \eqref{eq:pinskerbound} to increase Alice's output power while satisfying \eqref{eq:tvbound}. Inspired by recent work \cite{wang22isitcoverttd} that characterized covert communication over a quantum channel using quantum 
fidelity, we employ the classical analog of quantum fidelity to instead bound \eqref{eq:tvbound} \cite[Th.~15.1.2]{lehmann05stathyp4ed} via the Cauchy-Schwarz inequality:
\begin{align}
    p_e^{(w)}
    &\geq \frac{1}{2}-\frac{1}{\sqrt{2}}H(P_0^{n},P_1^n)\label{eq:cauchybound}
\end{align}
where $H(P_0^{n},P_1^n)\triangleq\sqrt{\frac{1}{2}\int_{\mathbb{C}^n}\dif^2\vec{w}\left(\sqrt{p_0^n(\vec{w})}-\sqrt{p_0^n(\vec{w})}\right)^{2}}$ is the Hellinger distance between $P_0^{n}$ and $P_1^{n}$.

Recall from  Section \ref{sec:system_design} that, when transmitting, Alice inputs QPSK symbol $x\in\mathcal{X}$ in each pulse slot. 
Then, Willie receives $\vec{w}_p(x)=h_{w}\vec{u}(x)+\vec{z}^{(w)}$ for $\vec{u}(x)=\left(e^{j\theta}\vec{c}_p,e^{j\theta}\vec{c}_q(x)\right)$, per the model in Fig.~\ref{fig:theoretical_channel}. Let $r_{p/q}=\frac{\|\vec{c}_p\|}{\|\vec{c}_q\|}$ be the ratio of magnitudes of the pilot and QPSK pulses, and $c_q=\|\vec{c}_q\|$ be the magnitude of the QPSK pulse.
Appendix \ref{ap:covertness} shows that we ensure $\delta$-covertness of the transmission scheme by setting
\begin{align}
        \alpha_n&\leq\frac{4\sqrt{2}\sigma_w^2}{a_w^2c_q^2}\sqrt{\frac{-\log\left(1-2\delta^2\right)}{n_p(1+r_{p/q}^4)}}\nonumber \\&=\left(\mathrm{SNR}\sqrt{1+r_{p/q}^4}\right)^{-1}\frac{4\sqrt{2}}{\sqrt{n_p}}\sqrt{-\log\left(1-2\delta^2\right)},\label{eq:anreq}
\end{align}
where $\mathrm{SNR}\triangleq \frac{h_{w}^2c_q^2}{\sigma^2_w}$ is Willie's received SNR and the penalty term $\sqrt{1+r_{p/q}^4}$ is due to inclusion of the pilot signal for channel phase estimate.
Combining 
\eqref{eq:anreq} with
$n_p=n/n_s$ and \eqref{eq:tp} yields the SRL scaling of covert throughput.

\subsection{Willie's Detectors}
\label{sec:willie_detectors}
The preceding analysis abstracts away detector structures available to Willie. We discuss them here, deferring their analysis to Appendices \ref{ap:test_statistic} and \ref{ap:willie_detector_errors}.
\subsubsection{Optimal Detector}
For known densities $p_0^n(\vec{w})$ and $p_1^n(\vec{w})$, a likelihood ratio test (LRT) is optimal and achieves $p_e^{(w)}$ in \eqref{eq:tvbound} with equality as $n\to\infty$ \cite[Th.~15.1.1]{lehmann05stathyp4ed}. 
LRT also minimizes $P_{\text{MD}}$ for a constant $P_{\text{FA}}$, per Neyman-Pearson lemma \cite[Th.~3.2.1]{lehmann05stathyp4ed}.
Here, $p_0^n(\vec{w})=\prod_{i=1}^{n_p}p^{(0)}\left(\vec{w}_i\right)$ and $p_1^n(\vec{w})=\prod_{i=1}^{n_p}p^{(1)}\left(\vec{w}_i\right)$, with $p^{(0)}\left(\vec{w}_i\right)$ and $p^{(1)}\left(\vec{w}_i\right)$ in \eqref{eq:prbpulsewillie0} and \eqref{eq:prbpulsewillie1}, where $\vec{w}_i$ are the observations of the $i^{\text{th}}$ pulse slot.
Willie accuses Alice if the LRT statistic $L(\vec{w})<0$, where
\begin{align}
    L(\vec{w})\triangleq\log\frac{p_0^n(\vec{w})}{p_1^n(\vec{w})}=\sum_{i=1}^{n_p}L_i\label{eq:LRT}
\end{align}
and $L_i=\log\frac{p^{(0)}(\vec{w}_i)}{p^{(1)}(\vec{w}_i)}$. Willie computes $L(\vec{w})$ from observations in $\vec{w}$ given his knowledge of Alice and Bob's system.
Unfortunately, Bessel functions in $p^{(1)}\left(\vec{w}_i\right)$ complicate the analysis of the resulting detector.
However, we expect the QPSK pulse magnitude $c_q$ to be small.
Employing the Taylor series expansion of $L(\vec{w})$ around $c_q=0$ in Appendix \ref{ap:test_statistic}, we simplify the LRT: Willie accuses Alice if the test statistic $S^{\text{(opt)}}(\vec{w})\geq\tau^{\text{(opt)}}$, where $S^{\text{(opt)}}(\vec{w})=\sum_{i=1}^{n_p}S_i^{\text{(opt)}}$,
\begin{align}S_i^{\text{(opt)}}&=\left|\left\langle \vec{c}_p,\vec{p}_i \right\rangle\right|^2+\left|\left\langle \vec{c}_q,\vec{q}_i \right\rangle\right|^2,\label{eq:Si}
\end{align}
$\vec{w}_i=(\vec{p}_i, \vec{q}_i)$ is the decomposition into pilot and QPSK segments, and $\tau^{\text{(opt)}}$ is the threshold discussed later.

\subsubsection{Power Detector}
Per \eqref{eq:Si}, Willie needs synchronization with Alice to use the optimal detector.  Radiometer offers a practical alternative: Willie accuses Alice if the total collected power $\left\| \vec{w}\right\|^2$ exceeds the threshold $\tau^{\text{(rad)}}$. For convenience, we define the radiometer test statistic $S^{\text{(rad)}}(\vec{w})=\sum_{i=1}^{n_p}S_i^{\text{(rad)}}$, where $S_i^{\text{(rad)}}=\left\| \vec{w}_i\right\|^2$ is in correspondence to \eqref{eq:Si}.

In Appendix \ref{ap:willie_detector_errors} we employ the Berry-Esseen theorem to analyze the probability of error achievable by these schemes, and derive the thresholds $\tau^{\text{(opt)}}$ and $\tau^{\text{(rad)}}$.
\revFinal{Furthermore, since the optimal detector achieves $p_e^{(w)}$ in \eqref{eq:tvbound}, we show in Appendix \ref{app:detector_asymptotics} how to improve $\alpha_n$ from \eqref{eq:anreq} to
\begin{align}
\alpha_n&\leq\left(\mathrm{SNR}\sqrt{1+r_{p/q}^4}\right)^{-1}\frac{4\sqrt{2}}{\sqrt{n_p}}\sqrt{\pi}\delta.\label{eq:anreqimproved}
\end{align}
However, we use more conservative \eqref{eq:anreq} in our experiments.\footnote{\revFinal{An anonymous referee's final review comment led to \eqref{eq:anreqimproved}. Regrettably, the publication deadline prevented us from using it in our experiments.}}
}

\section{Experiment Implementation}
\label{sec:implementation}
\subsection{System Configuration}
\label{sec:Experiment_Setup}

\begin{figure}
    \centering
    \includegraphics[width=1\linewidth]{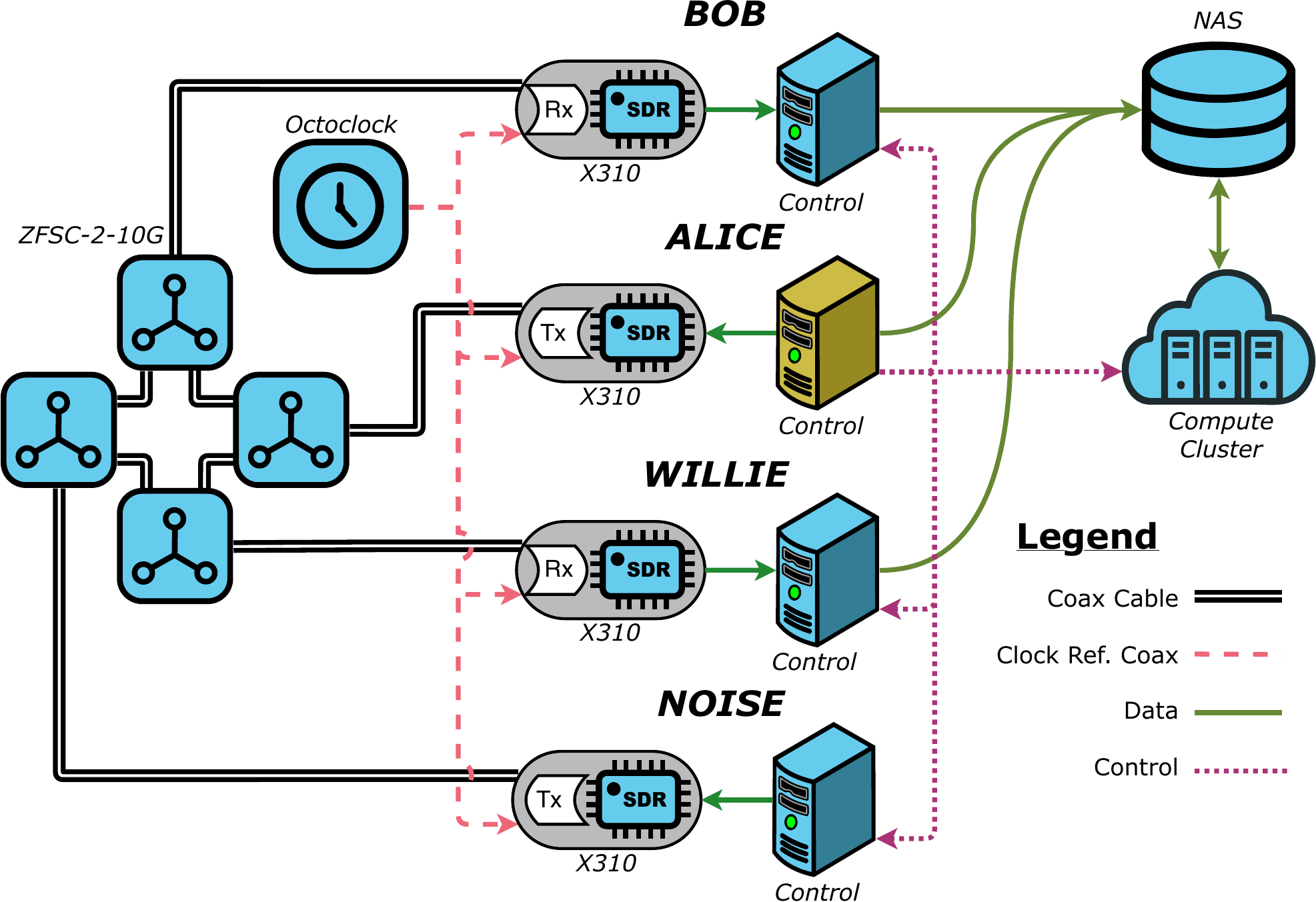}
    \caption{Covert communication experiment on COSMOS. Four Ettus USRP X310 radios -- Alice (Tx), Bob (Rx), Willie (warden), and a broadband noise source -- are linked by coaxial cables in a star topology using Mini-Circuits ZFSC-2-10G splitters/combiners. Tx-to-Rx and Tx-to-Tx path losses are 50 dB and 65 dB, respectively. All radios operate at \(f_c = 915\,\text{MHz}\) with Alice, Bob, and Willie using a DAC/ADC sampling rate of \(f_s = 12.5\times10^6\) samples/s and the noise generator $f_s = 22.5\times10^6$ samples/s. Alice, Bob, and Willie apply 0 dB Tx/Rx gain while the noise generator applies 15 dB gain. Each X310 connects over a 10 Gb/s SFP+ link to a dedicated control node (Intel Xeon E5-2640, 20 cores); Alice's node orchestrates the experiment via TCP messages to the other nodes and an eleven-node compute cluster of the same machines that performs real-time processing, while a 2 TB network-attached storage (NAS), mounted via NFS v4.2, provides a shared buffer.
    \label{fig:detailed_schematic}}
\end{figure}

We perform our experiments on COSMOS (formerly ORBIT), an open-access radio grid testbed \cite{raychaudhuri05orbit, orbit_grid}.
Our system is depicted in Fig.~\ref{fig:detailed_schematic}.
We connect four Ettus universal software radio peripheral (USRP) X310 SDRs, each fitted with a UBX 160 -- corresponding to transmitter Alice, receiver Bob, adversary Willie, and an AWGN generator -- into a shared-medium RF star network via coaxial cables.
This isolates our experiment from others on COSMOS, enables control over the environment, and allows repeatability.
We defer a wireless experiment with an organic noise source to future work.
Our radios are not power-calibrated. Therefore, we report measurements of signal amplitude and power in arbitrary units, respectively denoted as a.a.u.~and a.p.u.
The radios are positioned so that Bob and Willie experience approximately equal attenuation from both the AWGN generator and Alice, corresponding to an adverse scenario in which Willie is close to Bob.
We utilize a single 12.5 MHz channel centered at $f_c=915$ MHz. Both the transmitting (Alice and noise generator) and receiving (Bob and Willie) radios are offset-tuned by 25 MHz.

Each radio has a dedicated enhanced small form-factor pluggable (SFP+) cable connecting it to a high-bandwidth router, and the router to a dedicated control node on the COSMOS grid \cite{orbit_grid}.
SFP+ ports operate at 10 Gbps to support a maximum transmission unit (MTU) of 8 kB.
This prevents packet drops due to an internal radio buffer overflow during an experiment.
We note that even a single packet drop can cause catastrophic misalignment of the covert symbols during transmission, rendering an entire experimental trial useless. 
COSMOS provides a network \cite{orbit_grid} connecting the control nodes to network attached storage (NAS) and to other nodes forming a compute cluster for processing the collected data.

The radios' internal clocks are synchronized using an Ettus OctoClock, which provides low-jitter pulse-per-second (PPS) and 10 MHz reference signals, allowing the radios to maintain a constant phase offset between each other \cite{octoclock_datasheet}. 
We note that a centralized clock source is merely an experimental convenience.
In practice, Alice, Bob, and Willie can synchronize \emph{independently} to any stable clock source, such as an atomic clock, prior to transmission.\footnote{In a separate experiment in our lab at the University of Arizona, we confirmed this using a Stanford Research Systems FS725 10 MHz Rubidium Frequency Standard connected to Ettus USRP N210 radios.}

Alice's node generates fresh pre-shared secret and message vectors $\vec{t}$ and $\vec{x}_{2n_t}$ for each experimental trial.
Bob and Willie's radios continuously sample the channel, while Alice supplements the RF trial packet described in Section \ref{subsec:radio_comms} with TCP control messages over the SFP+ connections that mark its precise start and stop times.
At the end of each trial, control nodes write their data to the network-attached storage (NAS): Alice's node stores $\vec{t}$ and $\vec{x}_{2n_t}$, whereas Bob and Willie's nodes write their raw IQ samples, padded with brief pre- and post-buffers.
Alice's node then alerts the compute cluster, which analyzes the new trial and deletes its data on completion.
Acting as a rolling buffer, the NAS keeps heavy data traffic off the radio-control links and holds disk usage well below its 2 TB capacity.
The Ettus USRP library and USRP hardware driver (UHD), \cite{uhd_github, uhd_manual} are used to interact with the radios. 

\subsection{Alice's Spectral Footprint and SDR Non-idealities}
\label{sec:SDRnonidealities}
Section~\ref{subsec:covertness} addresses the fundamental limitations of implementing covert communication on digital transceivers.
Here, we identify the constraints specific to our SDR platform.

Our analysis assumes that Alice's signal energy is confined to a single $12.5$ MHz channel.  In practice, this is often violated: time-limited signals are unbounded in the frequency domain, and transmitter hardware non-idealities can generate strong spectral artifacts.
To characterize the emitted spectrum, we capture Alice's waveform with a Tektronix RSA5115B real-time spectrum analyzer positioned at Willie's location to record his observations exactly. The analyzer is centered at $915$ MHz, samples at $2\times10^8$ samples/s, and the external AWGN generator remains disabled throughout these spectral measurements. Alice transmits the shortest Gaussian pulses permitted by the system, $n_s^{(p)} = n_s^{(q)} = 6$, at an amplitude of $m_p = 0.12351232, m_q = 1.3313578$.

\begin{figure}[tb]
  \centering
  \includegraphics[width=0.9\linewidth]{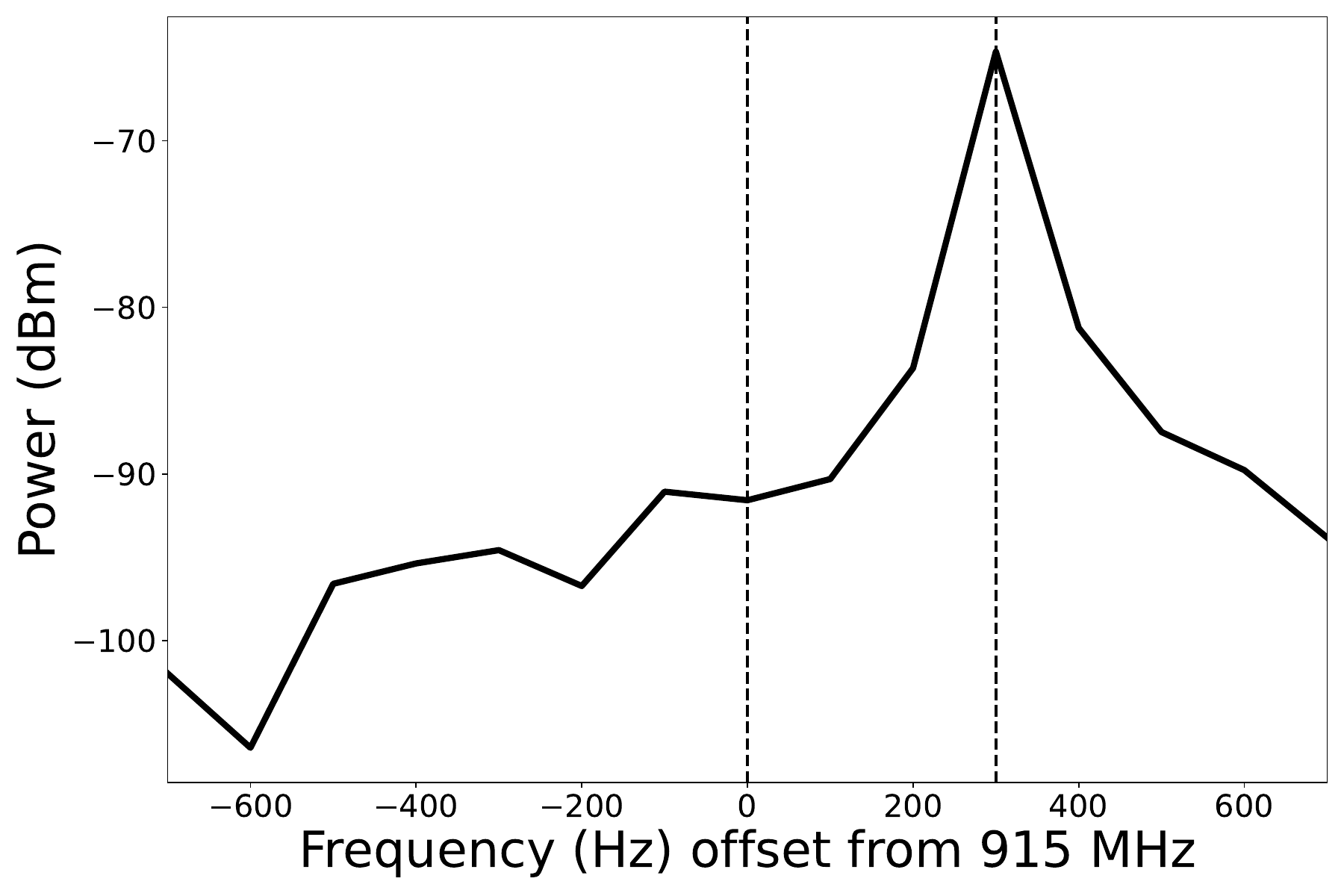}
  \caption{Measured spectrum at Willie's receiver. The Gaussian main lobe, expected to be centered at 915\,MHz, is displaced by approximately $+300$\,Hz, indicating a small carrier-frequency offset.}
  \label{fig:spectral_LO_offset}
\end{figure}

Fig.~\ref{fig:spectral_LO_offset} plots the spectrum obtained by integrating over many successive pulses; the Gaussian main lobe is shifted by roughly $+300$ Hz from the nominal 915 MHz. This offset exceeds the datasheet specification for a GPS-disciplined 10 MHz reference, which claims a frequency accuracy of $\pm25$ parts per billion (ppb), or $\pm23$ Hz at 915 MHz \cite{ettus_x310_spec}.

\begin{figure*}[!t]
  \centering
  \subfloat[Transmitted and received pulse pair (time domain)]{
    \includegraphics[width=0.48\textwidth]{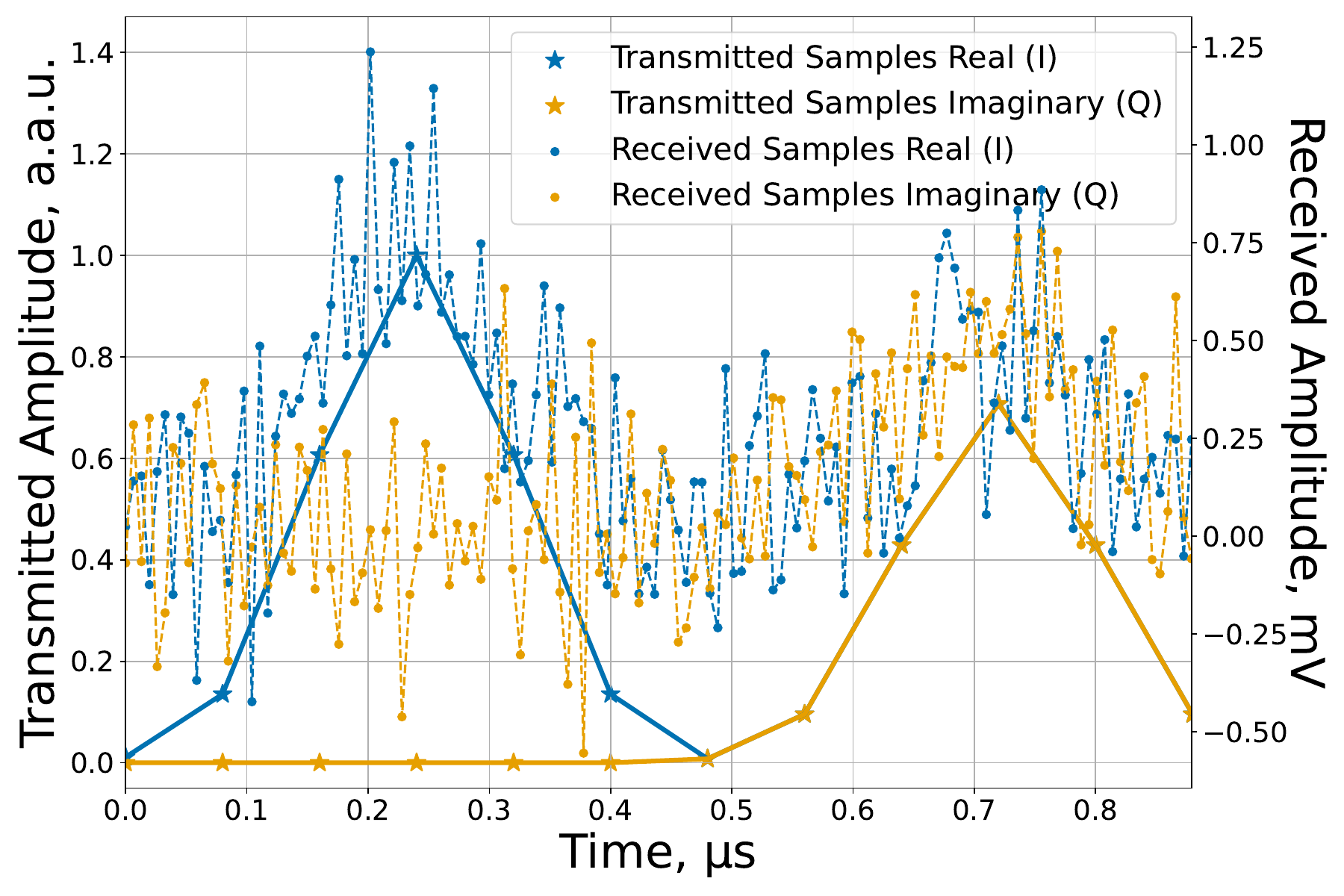}\label{fig:gaussian_single_pulse_iq}
  } \hfill
  \subfloat[Transmitted and received spectra]{
    \includegraphics[width=0.48\textwidth]{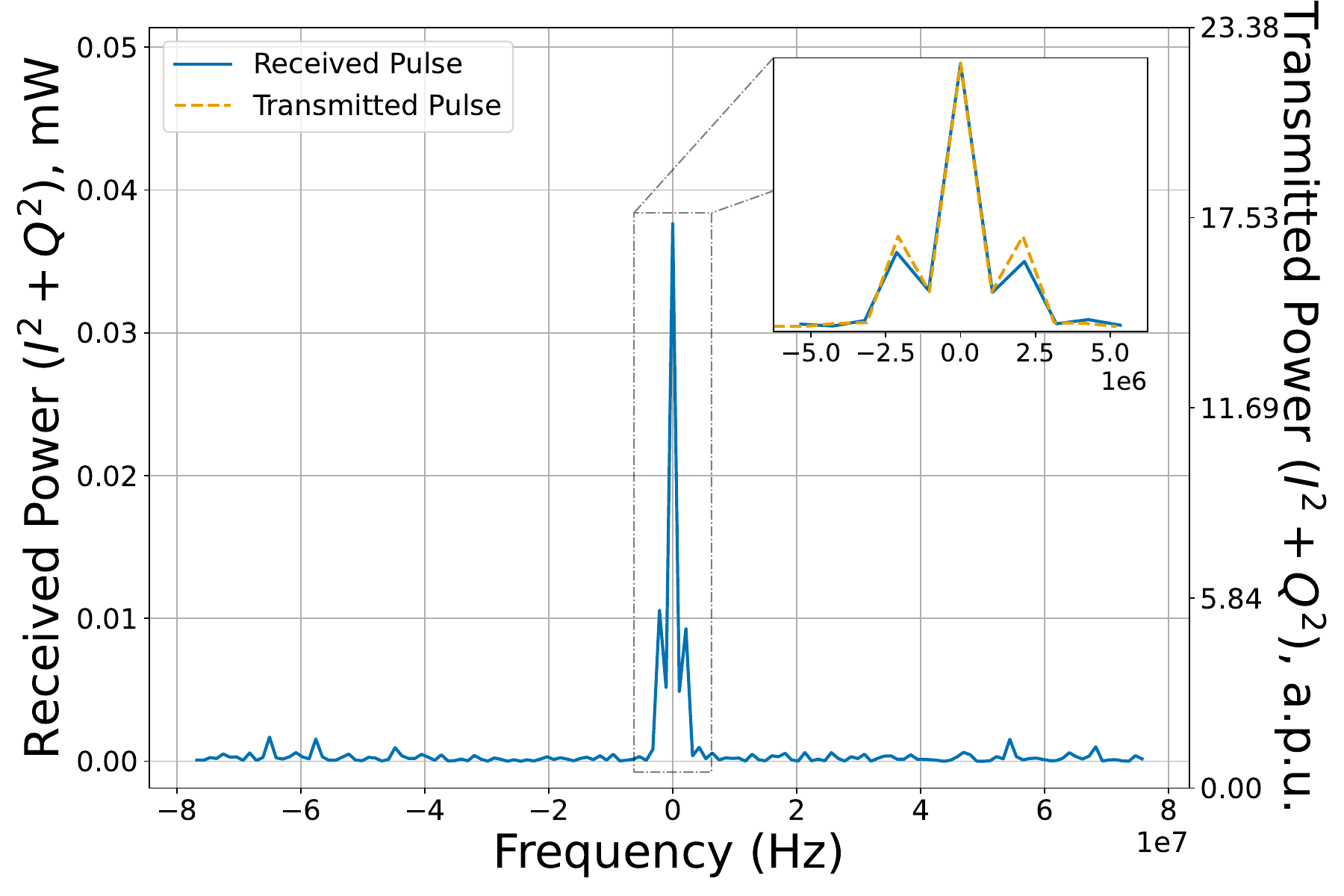}\label{fig:gaussian_single_pulse_spectrum}
  }
  \caption{Carrier-band verification of the 12-sample pulse pair transmitted by Alice at 0 dB gain. Each pair comprises a 6-sample Gaussian pilot followed immediately by a 6-sample Gaussian QPSK-data pulse. The left panel overlays the time-domain waveforms at the transmitter and at a spectrum analyzer placed at Willie, while the right panel shows the spectrum analyzer output, with insets overlaying the transmitted and received spectra in the carrier band. The amplitude and power transmitted by our uncalibrated radio are in arbitrary amplitude and power units (a.a.u.~and a.p.u.).}
  \label{fig:gaussian_single_pulse}
\end{figure*}

Fig.~\ref{fig:gaussian_single_pulse}\subref{fig:gaussian_single_pulse_iq} overlays the pilot-data pulse pair at the transmitter with the corresponding waveform captured by the spectrum analyzer; the traces align closely, confirming that the temporal shape is maintained.  Fig.~\ref{fig:gaussian_single_pulse}\subref{fig:gaussian_single_pulse_spectrum} presents the received spectrum over a wide frequency span and includes an inset that zooms to the carrier band observable at the transmitter. The transmit and receive spectra are overlaid in the inset and coincide nearly identically, confirming that the Gaussian envelope and its 99.9\% energy bandwidth are maintained. The full-span view further shows that essentially all energy is confined to this zoomed region; only a few narrow spikes appear in the immediate vicinity of the band edges, and with a single capture, it is unclear whether these are random artifacts or a repeatable spectral feature.

\begin{figure*}[!t]
  \centering
  \subfloat[Repeated pilot-data pulse pair]{
    \includegraphics[width=0.45\textwidth]{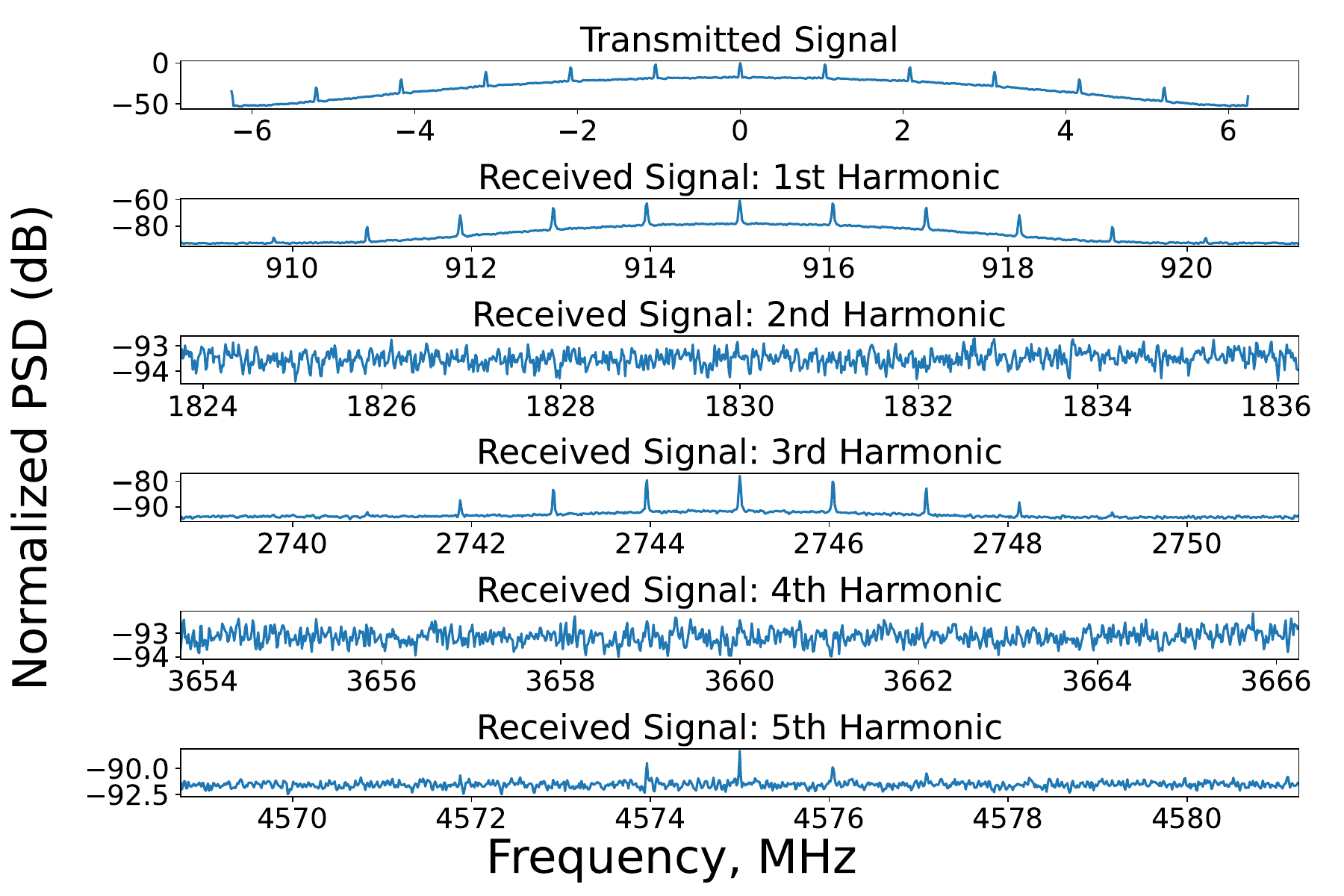}\label{fig:harmonics_tx_0dB}
  }\hfill
  \subfloat[All-zero waveform (silent mode)gain]{
    \includegraphics[width=0.45\textwidth]{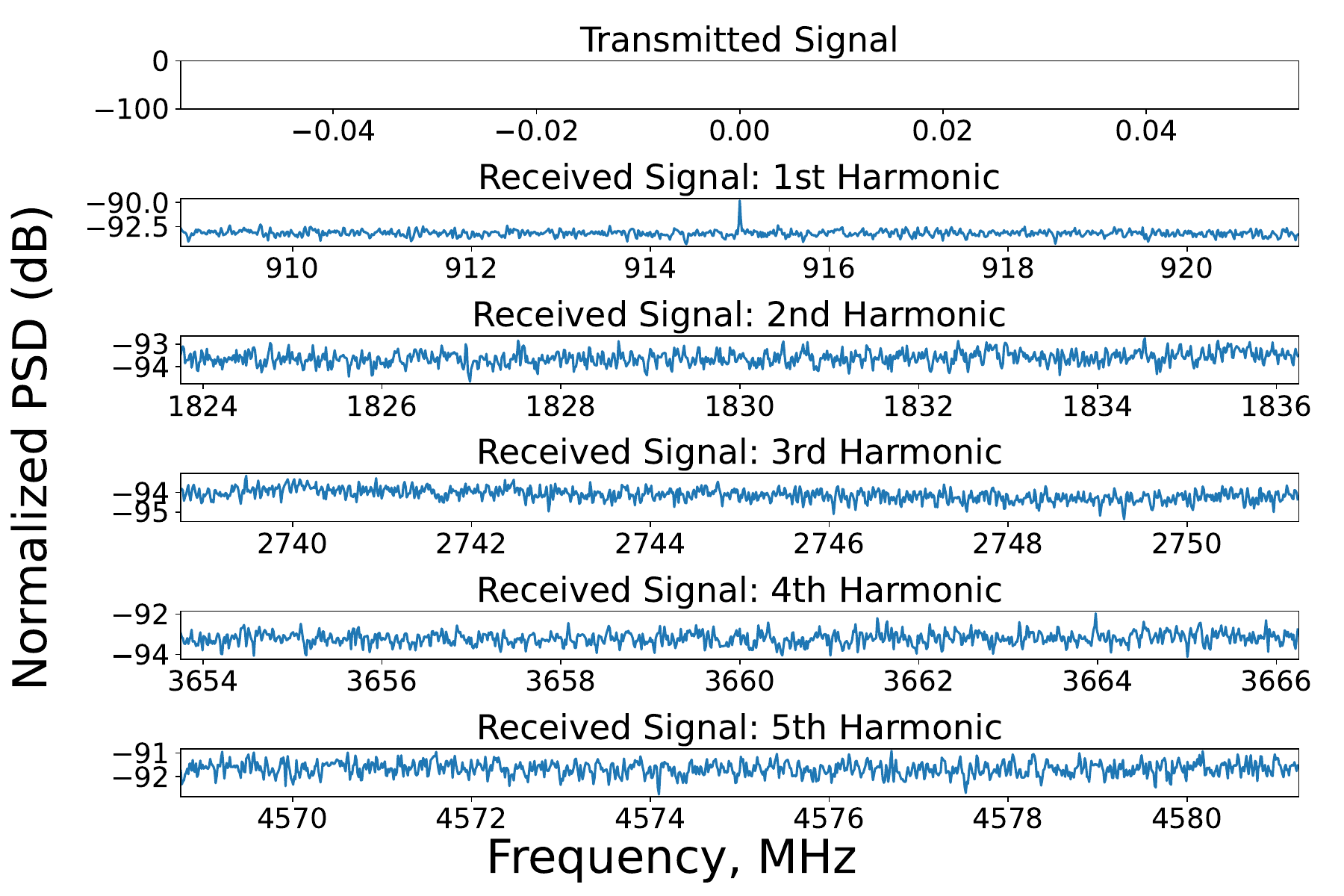}\label{fig:harmonics_silent_0dB}
  }\hfill
  \caption{Spectra at the first five harmonics, measured at Willie with Alice’s transmitter gain set to 0 dB. Left: when Alice continuously repeats the pilot-data pulse pair. Right: when the baseband input is forced to all zeros (silent mode). The ordinate scale is in dB relative to peak transmitted power.}
  \label{fig:harmonics_gain_compare}
\end{figure*}

Farther out in frequency, Fig.~\ref{fig:harmonics_gain_compare}\subref{fig:harmonics_tx_0dB} shows two pronounced spurs at the third and fifth harmonics of the carrier. Odd harmonics typically arise when a nominally sinusoidal signal develops a more square-like waveform, which in a transmitter can happen either through power-amplifier compression or through non-sinusoidal drive in the local oscillator (LO) chain. Neither explanation fits neatly here: the measurement is made at 0 dB hardware gain, well below the power amplifier’s compression point, and the UBX-160 LO path is followed by a 2.2 GHz low-pass section that should strongly attenuate components at 2.75 GHz and 4.58 GHz. The precise mechanism that leaves these third- and fifth-harmonic spurs above the noise floor, therefore, should be investigated.

To verify that Alice's silent mode is truly quiet, we transmit an all-zero waveform. Even with an all-zero signal, the spectrum in Fig.~\ref{fig:harmonics_gain_compare}\subref{fig:harmonics_silent_0dB} shows a small spur at the carrier and faint harmonics, behavior attributable to LO leakage. 

The two spectral artifacts most relevant to covertness at the settings used in our experiments are (i) the power that spills into odd harmonics and (ii) the residual carrier spur that appears when Alice is commanded to remain silent. The former could be mitigated by inserting an external low-pass filter at Alice's RF output, while the latter may be reduced through a careful power calibration of the X310 transmit chain. Making these improvements is challenging on shared-testbed hardware with limited on-site access. We will pursue this in the follow-on work. 

The radio circuitry is always on in our experiments. However, prior theoretical work and the analysis in Section \ref{sec:prerequisite} effectively assume that Alice's transmitter is completely powered off unless it is transmitting a pulse. In practice, power cycling a radio consumes significant time and may emit unwanted signals. That being said, we will investigate powering it off in the future, as the long intervals between sparse-coded transmissions may permit it. 
Furthermore, in some practical scenarios, $H_0$ can correspond to Alice's transmitter being always on but transmitting zero, as in our implementation.

Finally, LO power leaks into the baseband due to imperfect isolation between the LO and RF ports. When combined with small frequency offsets of the LO from the desired center frequency $f_c$, this leakage manifests as high-amplitude low-frequency noise. To mitigate this, we offset-tune our radios by 25 MHz, shifting the signal of interest away from this DC-centered noise.

\subsection{Pulse Optimization}
\label{subsec:hyperparam_opt}

To maximize the number of covert bits that can be reliably conveyed, we jointly tune four pulse-design parameters: the pilot-pulse length $n^{(p)}_{s}$, data-pulse length $n^{(q)}_{s}$, pilot-pulse magnitude $c_{p}$, and data-pulse magnitude $c_{q}$. We model the relationship between these parameters and the attainable covert bit count as a black-box objective, apply Bayesian optimization to efficiently navigate the parameter space, and use our hardware testbed to measure the covert bits delivered for each candidate configuration.
\begin{figure*}[t!]
\centering
\includegraphics[width=0.99\textwidth]{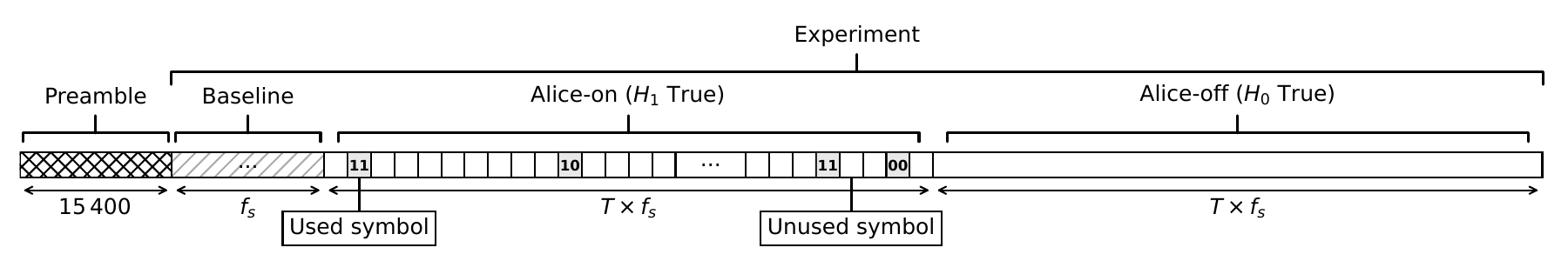}
\caption{Data-packet structure (durations in samples).  A 15\,400-sample preamble is created by concatenating five 13-bit Barker sequences, BPSK-modulating them, zero-padding each to 200 samples, and convolving the result with a 2\,401-sample root-raised-cosine filter. This is immediately followed by a one-second ($f_s$-sample) silence interval to measure the baseline noise level. The subsequent $T$-second, $T\times f_s$-sample segments encode the two parts of the experimental trial. In the Alice-on segment, each shaded pulse slot begins with a strictly positive, real-valued Gaussian pilot pulse (for phase reference), immediately followed by a Gaussian-shaped, QPSK-modulated data pulse carrying randomly generated information bits. The indices of the occupied slots are drawn uniformly at random and stored in $\vec{t}$ (see Section \ref{subsec:sysmodel}). In the subsequent Alice-off segment, she remains silent.}
\label{fig:data_packet}
\end{figure*}
Bayesian optimization iteratively explores the four-dimensional design space
$\vec{d} = \left(n^{(p)}_{s}, n^{(q)}_{s}, c_{p}, c_{q}\right)$ by maintaining a \emph{surrogate model} of the unknown mapping from $\vec{d}$ to attainable covert bit count. We leverage the Ax library in Python \cite{AxPlatform2025} to perform this optimization. The three steps that follow are the standard Bayesian optimization loop \cite[Ch.~31]{hennig2022probabilistic}; we restate them here.
\begin{itemize}
  \item \textbf{Propose}\,: a new configuration $\vec{d}_u$ is generated from the surrogate using all observations collected so far.
  \item \textbf{Measure}\,: using $\vec{d}_u$, the radio transmits a modified version of the data packet (Section~\ref{subsec:radio_comms}): every fifth pulse slot in the Alice-on segment carries a random bit, and the Alice-off segment is removed. The number of attainable covert bits is recorded over $100$ one-second trials. The average number of attainable covert bits for a given parameter set is $r = 2\,C_{\mathrm{bsc}}\,n_p\,\alpha_{n}$ bits over $n$ uses of the channel. Here, $n$ corresponds to the number of samples available in a one-second transmission. This yields a sample mean, $\bar{r}_u$, and a standard error, $s_{\bar{r}_u}$.
  \item \textbf{Update}\,: the tuple $(\vec{d}_u, \bar{r}_u, s_{\bar{r}_u})$ is appended to the data set and the surrogate is refitted before the next proposal.
\end{itemize}

After 200 iterations we use $\vec{d}^\star = \operatorname*{arg\,max}_{1\le l\le 200}\bar r_l,$ the configuration with the highest observed mean throughput.

We use a Gaussian-process surrogate, which is commonly adopted in Bayesian optimization. We briefly review it here. For a set of $l$ different input parameter configurations $\vec{d}_{1:l} = [\vec{d}_1, \vec{d}_2, ... \vec{d}_l]$, the model predicts $l$ corresponding covert throughput values sampled from the distribution $\mathcal{GP}\left(0,\,k(\vec{d}_{1:l},\vec{d}_{1:l})\right)$ \cite{williams2006gaussian}, where $k$ is a covariance function, which is the Mat\'{e}rn-5/2 function \cite[Ch.~4]{williams2006gaussian} here. This model can also be used to predict the throughput given priors, i.e., given $u-1$ tuples $(\vec{d}, \bar{r}, s_{\bar{r}})$, the model predicts the throughput $\hat{r}_u$ for some new parameter set $\vec{d}_u$. This prediction is distributed as $\hat{r}_u \mid \bar{r}_{1:(u-1)},\, \vec{d}_{1:u},\, s_{\bar{r}_{1:(u-1)}} \sim \mathcal{N}(\mu_u, \sigma_u^2)$, where $\mu_u$ and $\sigma_u$ can be directly calculated. At iteration $u$, configuration $d_u$ that maximizes \emph{expected improvement} $\mathbb{E}[\hat{r}_u' \mid \hat{r}_u' > \bar{r}^{\star}_{1:(u-1)}]$ for a candidate parameter configuration $d_u'$ is proposed for measurement, where $\hat{r}^{\star}_{1:(u-1)}$ is the highest measured throughput for the first $u-1$ trials, i.e., $r^{\star}_{1:(u-1)} = \max_{1\leq v \leq (u-1)} \bar{r}_v$.

The resulting optimum $(n^{(p)}_{s}=26, n^{(q)}_{s}=34, c_{p}=2.9765, c_{q}=3.521)$ is adopted for the experiments that follow.

\subsection{Transmission Structure for Experimental Trials} 
\label{subsec:radio_comms}

In each experimental trial, Alice transmits the four-segment packet shown in Fig.~\ref{fig:data_packet}: a non-covert \emph{preamble}, a baseline-estimation gap, an \emph{Alice-on} segment, and an \emph{Alice-off} (noise-only) segment. An AWGN generator operates continuously throughout the experiment. The Alice-on and Alice-off segments last $T$ seconds each, while the preamble marks the start of each trial and facilitates finding Alice’s transmission in Bob's and Willie's traces.

The preamble consists of the 13-bit Barker sequence repeated five times (65 bits in total). Each bit is BPSK-modulated and occupies 200 samples; the first sample carries the modulated symbol, followed by 199 zero-padding samples. The resulting 13\,000-sample sequence is then pulse-shaped with a root-raised-cosine (RRC) filter with roll-off factor $\beta = 0.35$, spanning 12 symbols and represented digitally by $12\times200 + 1 = 2401$ taps. After filtering, the preamble length is $65\times200 + (2401 - 1) = 15,400$ samples. The baseline-estimation gap is one second ($f_s$-samples) long, and is used to monitor noise power level throughout the experiment.

During the Alice-on segment, Alice embeds her covert message using the pilot- and data-pulse pairs defined in Section \ref{subsec:covert_pulse_design}, transmitting only in the pulse slots indexed by $\vec{t}$. We evaluate covert throughput and Bob's decoding reliability, as well as Willie's detection performance under hypothesis $H_1$, using the Alice-on segment. The Alice-off segment, which contains only AWGN, is used to characterize Willie's detector under the null hypothesis $H_0$.

\subsection{Experiment design}
\label{subsec:exp_design}
We select eight logarithmically-spaced values of Alice's transmission duration $T\in[10^{-0.3}, 10^{0.75}]$ s. 
For each value of $T$, we conduct $N=\N$ independent trials using distinct transmission packets described in Section \ref{subsec:radio_comms}.
Before we begin our experiment, we estimate Willie's SNR using a $20$ two-second calibration transmission with a modified version of the data packet from Section~\ref{subsec:radio_comms}: every fifth pulse slot in the Alice-on segment is used to transmit a random bit, and the Alice-off segment is deleted. The SNR is computed from the mean power values of the pilot, pulse, and noise, and the full estimator is detailed in Appendix~\ref{ap:snr}. We use the SNR estimate to compute 
$\zeta=\alpha_n\sqrt{n}=$, where $n=n_pn_s$ and $\alpha_n$ is in \eqref{eq:anreq}.
We set $\delta=0.07$ and use the result to compute $\alpha_n$ for each value of $T$.
This models Alice following the SRL.
We also evaluate two scenarios when Alice carelessly does not follow the SRL, where $\alpha_n=\{6.6\times 10^{-4},3.31\times^{10^{-3}}\}$, the first corresponding to $\alpha_n$ for the SRL at $T=10^{-0.3}$ s.
We use these three values of $\alpha_n$ to generate random vectors $\vec{t}$ with pulse locations used by Alice to transmit for each trial.

\begin{figure}[t]
\centering
    \includegraphics[width=0.98\columnwidth]{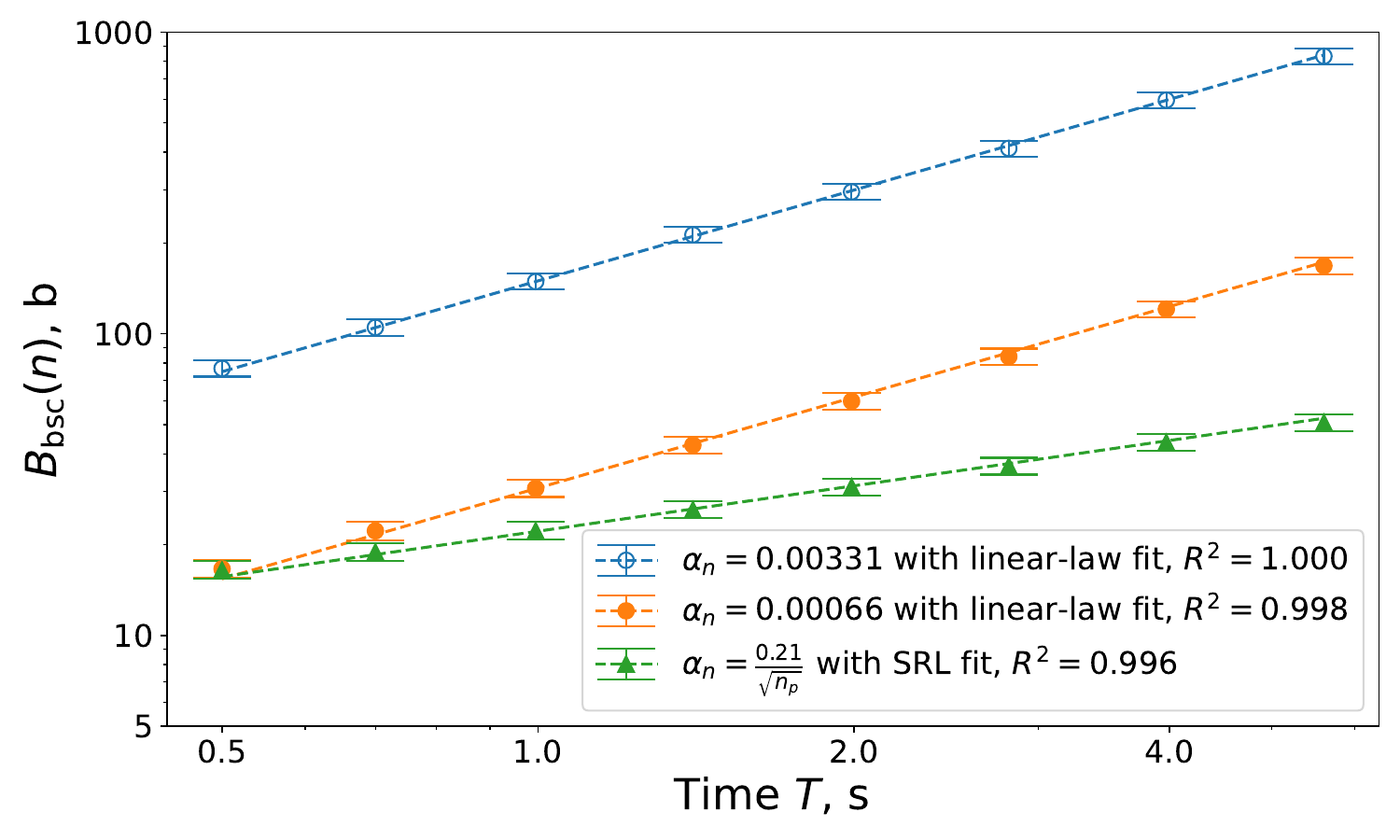}
\caption{Bob's receiver performance. We plot the total number $B_{\mathrm{bsc}}(n)$ of reliably-decodable bits for $n=f_s T$ channels uses vs.~transmission duration $T$ on a log-log scale, with 95\% confidence intervals as error bars give $N=\N$ trials per datapoint. $R^2$ is the coefficient of determination. 
}
\label{fig:log_log}
\end{figure}

\section{Results}
\label{sec:results}
\subsection{Covert Throughput}\label{sec:Bob_results}
Fig.~\ref{fig:log_log} plots Bob's receiver's performance vs.~transmission duration $T$.
Bob uses $\vec{t}$ to estimate bits only at the pulse locations Alice uses for transmission, per the sparse coding described in Section \ref{subsec:sysmodel}.
We report the estimates of the total number of transmissible covert bits $B_{\rm bsc}(n)$ by averaging over $N=\N$ trials using the equality in \eqref{eq:tp}.  Per Section \ref{subsec:exp_design}, Alice either follows the SRL or does not. We fit lines with slopes of either one-half or unity to the corresponding log-log plot results.
We calculate the coefficient of determination $R^2$, which is the proportion of the variation in $B_{\rm bsc}(n)$ explained by the square-root or linear dependence on $T$ \cite[Ch.~1.3]{draper98regression}. Our values of $R^2>0.99$ indicate the scaling results that we anticipate. While careless Alice can transmit significantly more bits, this makes her detectable, as discussed next.

\subsection{Detectability}
\label{sec:detectability}
We employ two methods to assess the detectability of Alice's transmission: directly by using a constant false-alarm rate (CFAR) strategy, and by evaluating the probability of error $p_e{(w)}$ in \eqref{eq:pew}.
For both approaches, we use the optimal detector and the radiometer discussed in Section \ref{sec:willie_detectors}.
We consider a scenario where the noise power level is static, which favors Willie.
We leave the extension to a dynamic environment, which could significantly favor Alice \cite{sobers17jammer}, for future work.

\subsubsection{CFAR Strategy}
\begin{figure*}[!th]
\centering
\subfloat[Optimal Detector]{%
\includegraphics[width=0.99\columnwidth]{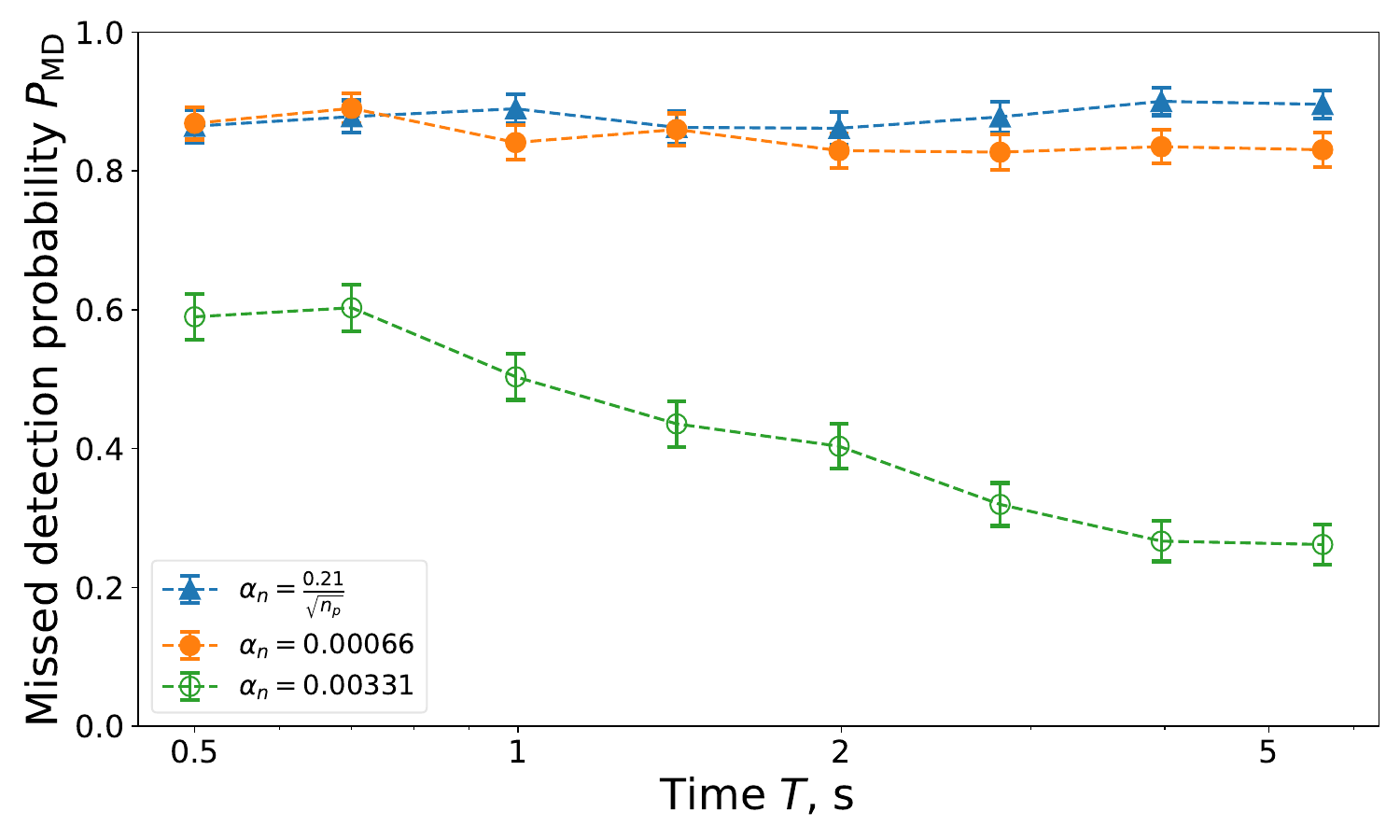}\label{fig:opt_perf_cfar}}
\hfill
\subfloat[Radiometer]{%
    \includegraphics[width=0.99\columnwidth]{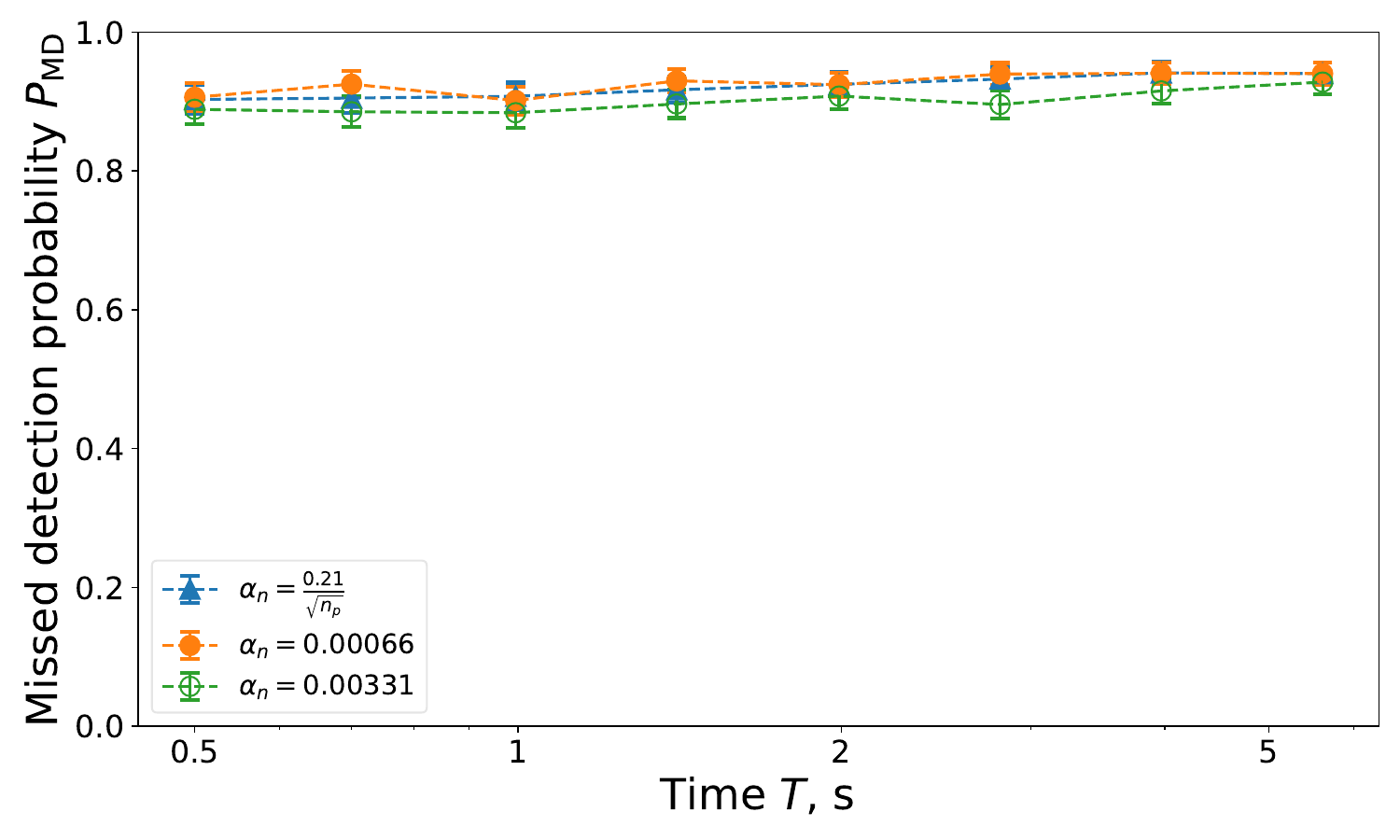}%
    \label{fig:rad_perf_cfar}%
}
\caption{Comparison of missed detection probability $p_{\text{MD}}$ under CFAR $P_{\text{FA}}^\star=0.1$ vs.~transmission duration $T$ (log scale) for \protect\subref{fig:opt_perf_cfar} the optimal detector and \protect\subref{fig:rad_perf_cfar} the radiometer. Results are averaged over $N=\N$ trials with 95\% confidence intervals shown.}
\label{fig:log_cfar}
\end{figure*}

Suppose Willie has a fixed false alarm budget $P_{\text{FA}}^\star$ and can collect detector data when Alice is absent.
He then estimates a CFAR threshold on his test statistic.
We implement this as follows: for a each value of $T$, and each detector in Section \ref{sec:willie_detectors}, we find a threshold $\tau^\star$ on a test statistic $S$ such that the number of trials where $S>\tau^\star$ given $H_0$ (i.e., the false alarm events) is  $P_{\text{FA}}^\star N$. Thus, $\tau^\star$ corresponds to a threshold that yields empirical false alarm probability $P_{\text{FA}}=P_{\text{FA}}^\star$.
We then report the corresponding fraction of trials when $S\leq \tau^\star$ given $H_1$, which estimates the probability of missed detection $P_{\text{MD}}$.
We note that our estimates of $P_{\text{MD}}$ and its 95\% confidence intervals depend on $\tau^\star$. 
It, in turn, depends on the measured noise power level, which is random.
However, tight control of our environment and a large number of trials used to calculate $\tau^\star$ render it nearly deterministic.
Fig.~\ref{fig:log_cfar} presents the results for $P_{\text{FA}}^\star=0.1$.
We compare the detectors at the end of this section.

\subsubsection{Evaluation of \texorpdfstring{$p_e^{(w)}$}{p\_e\^(w)}}
\begin{figure*}[!th]
\centering
\subfloat[Optimal Detector]{%
\includegraphics[width=0.99\columnwidth]{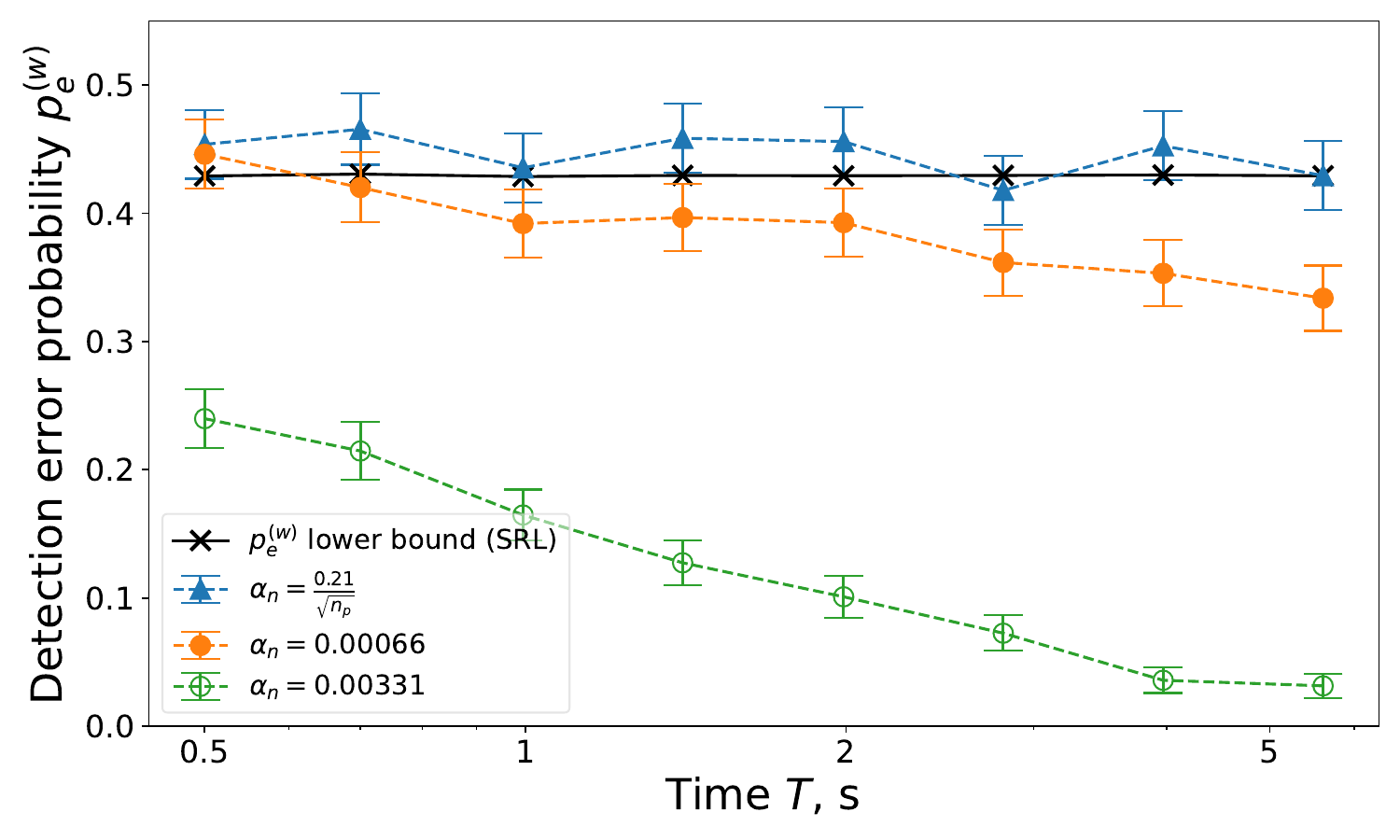}\label{fig:opt_perf}}
\hfill
\subfloat[Radiometer]{%
    \includegraphics[width=0.99\columnwidth]{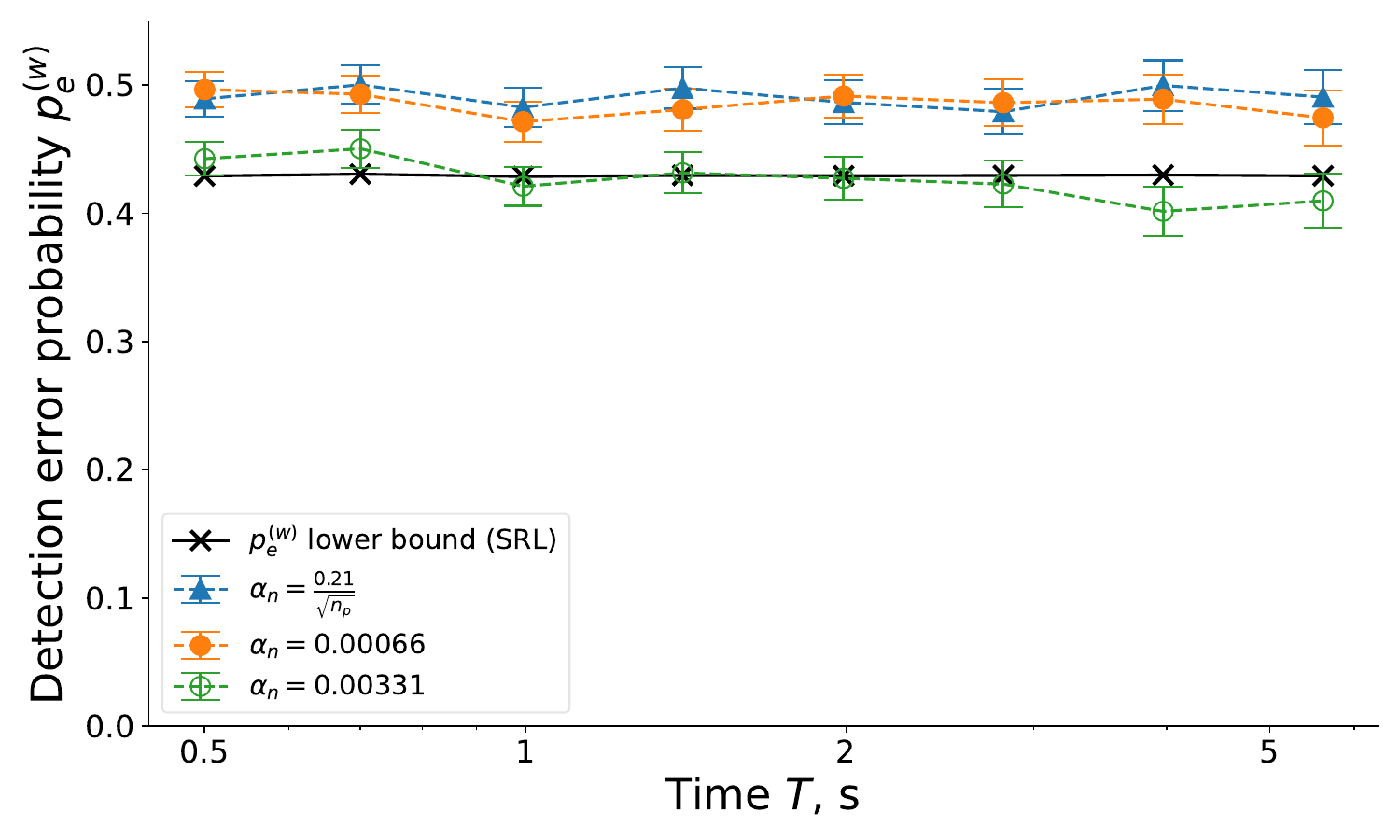}%
    \label{fig:rad_perf}%
}
\caption{Comparison of detection error probability $p_e^{(w)}$ versus transmission duration $T$ (log scale) for \protect\subref{fig:opt_perf} the optimal detector and \protect\subref{fig:rad_perf} the radiometer. Results are averaged over $N=\N$ trials with 95\% confidence intervals shown. The analytical lower bound from \eqref{eq:cauchybound} is calculated using the power from the same trials, but its error bars are negligible and omitted for clarity.}
\label{fig:log}
\end{figure*}

Fig.~\ref{fig:log} presents our estimates of Willie's probability of error $p_e^{(w)}$ in \eqref{eq:pew}.
First, we obtain the lower bound on $p_e^{(w)}$ when Alice follows the SRL by computing the upper bound on the Hellinger distance between the probability distributions of his received signals under each hypothesis in Appendix \ref{ap:covertness}. 
This is, effectively, a lower bound \eqref{eq:cauchybound} on a lower bound \eqref{eq:tvbound}.
We estimate the SNR for each trial as described in Appendix \ref{ap:snr}, calculate the bound, and plot the average over $N=\N$ trials on both subfigures in Fig.~\ref{fig:log}.
The correspondence to the covertness requirement $p_e^{(w)}\geq\frac{1}{2}-\delta=0.43$
indicates that our system operates as expected and that the noise power level and loss are stable throughout our experiments. 

We also investigate the performance of the optimal detector and radiometer described in Section \ref{sec:willie_detectors}.
For each trial, we estimate the threshold $\frac{\sqrt{n_p}\Delta\mu}{\sigma_0+\sigma_1}$ for each detector using Willie's observations as described in Appendix \ref{ap:snr} and calculate the estimate for $p_{e,\text{det}}^{(w)}$ using \eqref{eq:pedet}.
This corresponds to Willie implementing the detectors using information on the squared magnitude $a_w^2$ of the gain on his channel from Alice (which we assume he has), as well as measurements of the background noise $\sigma_w^2$.
We note that the Berry-Esseen estimation errors $\epsilon_{n_p,0}$ and $\epsilon_{n_p,1}$ are negligible due to $n_p$ being large.
We report the averages for each detector over $N=\N$ trials.
\subsubsection{Detector Comparison}
Both strategies show that Willie's probability of error remains constant when Alice follows the SRL.
When Alice is careless, the optimal detector's probability of error decays fairly rapidly with time $T$.  
On the other hand, the radiometer does not perform well in our experiments, even when $\alpha_n$ is a constant (corresponding to careless Alice).\footnote{In a separate experiment we used a larger constant $\alpha_n$, and confirmed that radiometer detects careless Alice.}
Nevertheless, this demonstrates how Alice can exploit shortcomings in Willie's detector design to improve her covert communication capabilities.

\section{Conclusion, Discussion, and Future Work} \label{sec:discussion}

We demonstrate the first implementation of SRL-based covert communication in the RF domain.
We employ SDRs, which are not specifically designed for covert communication. We have to address significant challenges:
\begin{itemize}
    \item \emph{Dynamic range and ADC/DAC granularity limitations}: The UBX-160 daughterboards use a 16-bit DAC for transmission and a 14-bit ADC for reception. While Alice can generate high-resolution low-power pulses, Bob's lower ADC precision limits his ability to distinguish weak signals from noise. Ensuring pulse detection after digitization requires minimal transmission power, which necessitates sparse coding. This, in turn, needs tight time synchronization, which we demonstrate.
    \item \emph{Time and frequency synchronization}: The oven-controlled crystal oscillators (OCXOs) in USRP X310s lack the frequency accuracy and stability needed to decode sparsely transmitted covert symbols.  Furthermore, Bob has to know precisely when Alice begins transmitting. We use a common reference (OctoClock) and a non-covert preamble; however, in practice, GPS or a highly stable time source, such as an atomic clock, can be used for both disciplining the local oscillators and for timing information. We will employ these in follow-on experiments. 
    \item \emph{Data volume:} Our capture for $N=\N$ trials amounts to approximately $72$ h of baseband data at a sampling rate of $f_s = 12.5$ MHz, using 64-bit in-phase and quadrature samples for each receiver (Bob and Willie). This produces roughly $10$ TB of data.
    Although real-time processing is performed on the compute cluster, both computational and storage demands must be reduced for practical deployment.
    \item \emph{Continuous noise injection}: Maintaining an always-on, spectrally flat artificial noise floor is limited by the maximum DAC sampling rate the system can sustain: sampling too fast can overwhelm the host CPU and lead to buffer under-flows. In the future, we will explore using multiple radios to emulate noise and more natural noise sources by experimenting ``in the wild.''
\end{itemize}

Indeed, our proof-of-concept experiment raises many theoretical and experimental research questions.
In the short term, we plan to address some of them by improving our COSMOS-based design as follows:
\begin{itemize}
    \item We will add control over path loss to study the impact of channel dynamics;
    \item \revFinal{We will look for an optimal constellation satisfying the covertness criteria in \eqref{eq:tvbound} and \eqref{eq:cauchybound}. There is evidence for this being QPSK, as it is optimal for complex-valued circularly-symmetric AWGN under the criterion in \eqref{eq:pinskerbound} (by extending the BPSK case for real-valued AWGN channels in \cite{bash13squarerootjsacnonote, bloch15covert, wang15covert}).} This is also actively investigated in quantum information theory \cite{wang22isitcoverttd}. 
\end{itemize}
Additionally, the AWGN channel model provides only a zeroth-order approximation to practical RF channels.
Thus, SRL-based covert communication needs to be validated in a more realistic, dynamic environment.
We plan on evolving our system to be independent rather than centrally controlled.
This would allow not only experimentation ``in the wild'' but also exploration of covert networks.
Furthermore, we plan to relax the assumption on the adversary's capabilities and study the impact of, e.g., the lack of precise knowledge of the timing of transmission and the pulse shape used.

\section*{Acknowledgment}
The authors are grateful to Ivan Seskar and the COSMOS staff for setting up the radios on COSMOS and answering our many questions.
The authors thank Loukas Lazos, Ming Li, Jingcheng Li, Ziqi Xu, Samuel H.~Knarr, and Timothy C.~Burt for valuable advice and for sharing the equipment.
The authors also acknowledge helpful discussions with Mark J.~Meisner, Jaim Bucay, Dennis L.~Goeckel, Donald F.~Towsley, Matthieu R.~Bloch, Matthew Arcarese, and Robert J.~McGurrin.
\revFinal{Finally, the authors thank the anonymous referees for their suggestions, which substantially improved the manuscript.}
\appendices

\section{Covertness Criterion Analysis}\label{ap:covertness}

Since Willie knows the start time, the duration of Alice's transmission, and details of her system from Section \ref{subsec:sysmodel}, he collects $n$ observations corresponding to the total number of channel uses available to Alice.
Furthermore, Alice randomizes the per-pulse phase $\theta$ since her pilot-aided QPSK scheme allows Bob to recover phase information on receive without further channel estimation. 
These $n$ observations are divided into $n_p$ pulse slots 
of $n_s=n^{(p)}_{s}+n^{(q)}_{s}$ observations each, corresponding to pulses containing $n^{(p)}_{s}$ (resp.~$n^{(q)}_{s}$) samples of pilot signal (resp.~QPSK symbol), as described in Section \ref{subsec:sysmodel}.
Each density function $p^{(h)}(\vec{w}_i)$ with segment $\vec{w}_i\in\mathbb{C}^{n_s}$ is indexed according to the true hypothesis $H_h$, $h\in\{0,1\}$ and pulse location $i=1,\ldots,n_p$. For hypothesis $H_1$ when Alice employs a pilot signal, we decompose $\vec{w}_i=\left(\vec{p}_i,\vec{q}_i\right)$ for pilot samples $\vec{p}_i\in\mathbb{C}^{n^{(p)}_{s}}$ and QPSK samples $\vec{q}_i\in\mathbb{C}^{n^{(q)}_{s}}$, respectively. Denote by $\psi(\vec{w};\vec{\mu}, \mathbf{\Gamma},\mathbf{C})$ the multi-dimensional complex Gaussian density function with mean vector $\vec{\mu}=E\left[\vec{W}\right]$, covariance matrix $\mathbf{\Gamma}=E\left[(\vec{W}-\vec{\mu})(\vec{W}-\vec{\mu})^\dagger\right]$, and relation matrix $\mathbf{C}=E\left[(\vec{W}-\vec{\mu})(\vec{W}-\vec{\mu})^T\right]$, defined as 
\begin{align}
\psi(\vec{w};\vec{\mu}, \mathbf{\Gamma},\mathbf{C})={\frac {\exp \!\left(-{\frac {1}{2}}{\begin{bmatrix}\vec{w}-\vec{\mu} \\{\overline {\vec{w}}}-{\overline {\vec{\mu} }}\end{bmatrix}}^{\dagger }{\begin{bmatrix}\mathbf{\Gamma} &\mathbf{C}\\{\overline {\mathbf{C}}}&{\overline {\mathbf{\Gamma}}}\end{bmatrix}}^{\!\!-1}\!{\begin{bmatrix}\vec{w}-\vec{\mu} \\{\overline {\vec{w}}}-{\overline {\vec{\mu} }}\end{bmatrix}}\right)}{\pi ^{n}{\sqrt {\det(\mathbf{\Gamma} )\det(\overline{\mathbf{\Gamma}}-\mathbf{C}^\dagger \mathbf{\Gamma}^{-1}\mathbf{C})}}}},
\end{align}
where $\overline{\mathbf{A}}$ denotes the complex conjugate of matrix $\mathbf{A}$.
Since our noise is i.i.d.~and induces circularly symmetric complex Gaussian distributions, $\mathbf{C}=\mathbf{0}$ and we drop it for brevity.
Under $H_0$, Alice does not transmit and Willie observes AWGN. Thus, segments are i.i.d.~with the density function:
\begin{align}
p^{(0)}\left(\vec{w}_i\right)&=\psi\left(\vec{w}_i;\vec{0}, 2\sigma^2_w \mathbf{I}_{n_s}\right).\label{eq:prbpulsewillie0}
\end{align}
Alice transmits under hypothesis $H_1$.
Since Willie does not have $\vec{t}$ and $\vec{s}$, segments are i.i.d.~with the density function: 
\begin{align}
p^{(1)}\left(\vec{w}_i\right)\nonumber&=\int_{0}^{2\pi}\dif\theta\left((1-\alpha_n)\psi\left(\vec{w}_i;\vec{0},2 \sigma^2_w \mathbf{I}_{n_s}\right) \right.\nonumber\\&\phantom{=\int==}\left.+ \frac{\alpha_n}{4}\sum_{x=0}^3\psi\left(\vec{w}_i;e^{j\theta}\vec{\mu}(x), 2\sigma^2_w \mathbf{I}_{n_s}\right)\right) \nonumber\\
&=(1-\alpha_n)\psi\left(\vec{w}_i;\vec{0}, 2\sigma^2_w \mathbf{I}_{n_s}\right) \nonumber\\ 
&\phantom{=}+ \frac{\alpha_n}{4}\sum_{x=0}^3\int_{0}^{2\pi}\dif\theta\psi\left(\vec{p}_i;e^{j\theta}h_{w}\vec{c}_p , 2\sigma^2_w 
\nonumber\mathbf{I}_{n_s}\right)\nonumber\\&\phantom{=+ \frac{\alpha_n}{4}\sum_{x=0}^3}\times\psi\left(\vec{q}_i;e^{j\left( \theta + \pi\frac{2x-1}{4}\right)}h_{w}\vec{c}_q , 2\sigma^2_w 
\mathbf{I}_{n_s}\right)\nonumber\\
&=\nonumber(1-\alpha_n)\psi\left(\vec{w}_i;\vec{0}, 2\sigma^2_w \mathbf{I}_{n_s}\right)\nonumber \\ 
&\phantom{=}+ \frac{\alpha_n}{4}\psi\left(\vec{w}_i;\vec{0}, 2\sigma^2_w \mathbf{I}_{n_s}\right)e^{-a_w^2\frac{\|\vec{c}_p\|^2+\|\vec{c}_q\|^2}{2 \sigma_w^2}}\nonumber\\
&\phantom{=+}\times\sum_{x=0}^3 I_0\left(\frac{a_w}{\sigma_w^2}f_x\left(\vec{w}_i,\vec{c}_p, \vec{c}_q\right)\right),\label{eq:prbpulsewillie1}
\end{align}
where $I_0(\cdot)$ is the modified Bessel function of the first kind with order $0$, $a_w$ is the channel loss,
and 
\begin{align}
    f_k&\left(\vec{w}_i,\vec{c}_p, \vec{c}_q\right)\nonumber \\&=\left(\left(\left\langle\vec{c}_p,\Re(\vec{p}_i)\right\rangle+(-1)^{\lfloor k/2 \rfloor}\left\langle\vec{c}_q,\Re(\vec{q}_i)\right\rangle\right)^2 \nonumber\right. \\
           &\phantom{=+}\left.\vphantom{\left((-1)^{\lfloor k/2}\right)^2}+\left(\left\langle\vec{c}_p,\Im(\vec{p}_i)\right\rangle+(-1)^k\left\langle\vec{c}_q,\Im(\vec{q}_i)\right\rangle\right)^2 \right)^{1/2}
\end{align}
We seek to bound Willie's probability of detection error. Cauchy-Schwarz inequality implies $\mathcal{V}_T(P_0^n, P_1^n) \leq \sqrt{2}H(P_0^n, P_1^n)$, thus we analyze the Hellinger distance between our distributions. We have 
\begin{align}
    H^2&(P_0^n, P_1^n) \nonumber\\&= 1-\int_{\mathbb{C}^n}\dif ^2\vec{w}_i\sqrt{\prod_{i=1}^{n_p}p^{(0)}\left(\vec{w}_i\right)p^{(1)}\left(\vec{w}_i\right)}\\&=1-\prod_{i=1}^{n_p}\int_{\mathbb{C}^{n_s}}\dif^2\vec{w}_i\sqrt{p^{(0)}\left(\vec{w}_i\right)p^{(1)}\left(\vec{w}_i\right)}\label{eq:Fubini}
    \\
    &= 1-\left(\int_{\mathbb{C}^{n_s}}\dif^2\vec{w}_1\sqrt{p^{(0)}\left(\vec{w}_1\right)p^{(1)}\left(\vec{w}_1\right)}\right)^{n_p}\label{eq:hdist_indep}
\end{align}
The term inside the power of $n_p$ on the right hand side of \eqref{eq:hdist_indep} admits no known closed-form expression, so we take its Taylor series expansion.
Without loss of generality, substitute $\vec{c}_q\to c_q\vec{b}_q$ where $c_q=\|\vec{c}_q\|$ is the magnitude of the QPSK pulse and $\|\vec{b}_q\|=1$ is the unit vector with the same pulse shape as $\vec{c}_q$. Similarly, substitute $\vec{c}_p \to c_q r_{p/q}\vec{b}_p$ where $r_{p/q}=\frac{\|\vec{c}_p\|}{\|\vec{c}_q\|}$ is the ratio of magnitudes the pilot pulse and the QPSK pulse, and $\|\vec{b}_p\|=1$ is the unit vector with the same pulse shape as the $\vec{c}_p$. We Taylor expand the term inside the power of $n_p$ about $c_q=0$, and obtain:
\begin{align}
\int_{\mathbb{C}^{n_s}}&\dif^2\vec{w}_1\sqrt{p^{(0)}\left(\vec{w}_1\right)p^{(1)}\left(\vec{w}_1\right)}\nonumber\\&=1-\alpha_n^2\frac{1+r_{p/q}^4}{32\sigma_w^4}a_w^4c_q^4+\alpha_n^3\frac{1+r_{p/q}^6}{64\sigma_w^6}a_w^6c_q^6+o(c_q^6)
\end{align}
Taylor's theorem with remainder implies 
\begin{align}
     H^2(P_0^n, P_1^n)&\leq1-\left(1-\alpha_n^2\frac{1+r_{p/q}^4}{32\sigma_w^4}a_w^4c_q^4\right)^{n_p}.
\end{align}
We seek to bound $\alpha_n$ to ensure covertness. Using covertness requirement $\delta$ and \eqref{eq:cauchybound} implies
\begin{align}
    \delta&\geq\sqrt{\frac{1}{2}\left(1-\left(1-\alpha_n^2\frac{1+r_{p/q}^4}{32\sigma_w^4}a_w^4c_q^4\right)^{n_p}\right)}
\end{align}
Rearranging yields
\begin{align}
    \alpha_n&\leq\frac{4\sigma_w^2}{a_w^2c_q^2}\sqrt{\frac{2\left(1-(1-2\delta^2)^{1/n_p}\right)}{1+r_{p/q}^4}}.\label{eq:alpha_bound_np_exp}
\end{align}
Now, we substitute $1/n_p\to x^2$ and Taylor expand the right-hand side of \eqref{eq:alpha_bound_np_exp} with respect to $x=0$ as 
\begin{align}
    &\frac{4\sigma_w^2}{a_w^2c_q^2}\sqrt{\frac{2\left(1-(1-2\delta^2)^{x^2}\right)}{1+r_{p/q}^4}} \nonumber\\&\phantom{==}=\nonumber  \frac{4\sigma_w^2}{a_w^2c_q^2}\sqrt{\frac{2\log\left(\frac{1}{1-2\delta^2}\right)}{1+r_{p/q}^4}}x\\
    &\phantom{===}-\frac{\sigma_w^2}{a_w^2c_q^2}\sqrt{\frac{2\log\left(\frac{1}{1-2\delta^2}\right)}{1+r_{p/q}^4}}\log\left(\frac{1}{1-2\delta^2}\right)x^3+o(x^3).
\end{align}
Then, for $n_p$ large enough, Taylor's theorem with remainder and reverting the substitution yields the expression in \eqref{eq:anreq}. 

\section{Phase Estimation}\label{ap:phase_est}

In each pilot segment of the pulse, Bob receives the complex-valued vector  
\begin{align}
\vec{y}_p&=a_b\vec{c}_p e^{\im\theta_b}+\vec{z}^{(b)}_p,
\end{align}
where $\vec{z}^{\,(b)}_p\in\mathbb{C}^{n_s^{(p)}}$ is circularly-symmetric complex-valued AWGN afflicting the pilot segment.
Applying the pilot-pulse-shape filter yields:
\begin{align}
\langle \vec{c}_p,\vec{y}_p\rangle&=a_{b}e^{\im\theta_{b}}\|\vec{c}_p\|^2+\langle \vec{c}_p,\vec{z}^{(b)}_p\rangle.
\end{align}
The expected values of the in-phase and quadrature (IQ) components $p_I\triangleq \Re\left(\langle \vec{c}_p,\vec{y}_p\rangle\right)$ and $p_Q\triangleq \Im\left(\langle \vec{c}_p,\vec{y}_p\rangle\right)$ are:
\begin{align}
    \mathbb{E}\left[\Re\left(\langle \vec{c}_p,\vec{y}_p\rangle\right)\right] &=\mathbb{E}[p_I] = a_{b}\|\vec{c}_p\|^2\cos(\theta_{b})\\
    \mathbb{E}\left[\Im\left(\langle \vec{c}_p,\vec{y}_p\rangle\right)\right] &=\mathbb{E}[p_Q] = a_{b}\|\vec{c}_p\|^2\sin(\theta_{b}),
\end{align}
since AWGN is circularly-symmetric and has zero mean.
The estimate of phase $\theta_b$ is then $\hat{\theta}_b=\tan^{-1}\frac{p_I}{p_Q}$.
As $n_s^{(p)}$ increases, $p_I$ and $p_Q$ converge to their expected values, yielding a more accurate estimate of phase. Here, $n_s^{(p)}=26$ suffices.

\section{Derivation of Optimal Detector Test Statistic}\label{ap:test_statistic}
As in Appendix \ref{ap:covertness}, we substitute $\vec{c}_q\to c_q\vec{b}_q$ and $\vec{c}_p \to c_q r_{p/q}\vec{b}_p$ in the expression for $L_i$ in \eqref{eq:LRT} and Taylor expand it about $c_q=0$:
\begin{align}
    L_i &= \frac{\alpha_n}{4\sigma_w^4}\left(r_{p/q}^2\left(|\langle \vec{b}_p,\vec{p}_i \rangle|^2-2\sigma_w^2||\vec{b}_p||^2 \right) \right.\nonumber \\ &\phantom{=\frac{\alpha_n}{4\sigma_w^4}\big(-}\left.+|\langle \vec{b}_q,\vec{q}_i \rangle|^2 -2\sigma_w^2\|\vec{b}_q\|^2\right)c_q^2 +o\left(c_q^2\right).
\end{align}
Thus, for $c_q$ sufficiently small, the $i^{\text{th}}$ pulse-slot component $S_i^{\text{(opt)}}$ of Willie's sufficient statistic is in \eqref{eq:Si}.

\section{Willie's Detectors' Errors}
\label{ap:willie_detector_errors}
\subsection{Analysis}
Here, we derive bounds for the probability of error attainable by the optimal detector and radiometer, described in Section \ref{sec:willie_detectors}.
We use the Berry-Esseen theorem and assume non-informative priors. 
Both test statistics $S^{\text{(opt)}}(\vec{w})$ and $S^{\text{(rad)}}(\vec{w})$ have the form $S=\sum_{i=1}^{n_p}S_i$, where $(S_i)$ is a sequence of realizations of $n_p$ i.i.d.~random variables.
Let us standardize this sum using $H_0$ parameters:
\begin{align}
S_{n_p} = \frac{S - n_p\mu_0}{\sigma_0\sqrt{n_p}},
\end{align}
where $\mu_0$ and $\sigma_0^2$ are the mean and variance of $S_i$ under $H_0$. 
Similarly, let $\mu_1$ and $\sigma_1^2$ be the mean and variance of $S_i$ under $H_1$, noting that they may depend on $n_p$.
We provide the necessary moments of $S_i$ under $H_0$ and $H_1$ for both optimal detector and radiometer in Appendix \ref{ap:moments}. 

Let $F_{n_p,h}(x)$ be the cumulative distribution function (c.d.f.) of the random variable statically describing $S_{n_p}$ under $H_h$, $h\in\{0,1\}$. By the Berry-Esseen theorem \cite[Ch.~XVI.5]{feller71probtheory2},
\begin{align}
\sup_{x} |F_{n_p,k}(x) - \Phi(x)| \leq B_{n_p,k} \triangleq \frac{C \rho_k}{\sigma_{k}^3 \sqrt{n_p}}, \label{eq:berry-esseen}
\end{align}
where $\rho_k$ and $\sigma_{k}$ are the absolute third central moment and the standard deviation of $S_i$ under $H_k$, $\Phi(x)$ is the c.d.f.~of standard Gaussian distribution $\mathcal{N}(0,1)$, and $C<0.4748$ is the Berry-Esseen constant. 
Willie decides between $H_0$ and $H_1$ by comparing $S_{n_p}$ to a threshold $\tau$, per Section \ref{sec:willie_detectors}. The probability of error given $\tau$ is $p_e^{(w)}(\tau) = \frac{1}{2} \left(P(S_{n_p} > \tau | H_0) +  P(S_{n_p} \leq \tau | H_1)\right)$.
Applying \eqref{eq:berry-esseen}, we have $P(S_{n_p} > \tau | H_0) = Q(\tau) + \epsilon_{n_p,0}$, where the approximation error $\epsilon_{n_p,0}\in[-B_{n_p,0},B_{n_p,0}]$ and $Q(x)\triangleq\frac{1}{\sqrt{2\pi}}\int_x^\infty e^{-t^2/2}\dif t$ is the Gaussian error function. Under $H_1$, we transform the event $S_n \leq \tau$ to the standardized frame of $H_1$. Defining the effective argument $z(\tau) \triangleq \frac{\tau\sigma_0 - \sqrt{n_p}\Delta\mu}{\sigma_1}$, where $\Delta\mu = \mu_1-\mu_0$, we have $P(S_{n_p} \leq \tau | H_1) = Q\left(-z(\tau)\right) + \epsilon_{n_p,1}$, where $\epsilon_{n_p,1}\in[-B_{n_p,1},B_{n_p,1}]$.
Thus, $p_e^{(w)}(\tau) = \frac{1}{2} \left(Q(\tau)+Q\left(-z(\tau)\right)\right)+\frac{\epsilon_{n_p,0}+\epsilon_{n_p,1}}{2}$.
The sum of $Q$-functions is minimized by symmetric arguments: $\tau=-z(\tau)=\frac{\sqrt{n_p}\Delta\mu}{\sigma_0+\sigma_1}$. Thus, the minimum error probability achievable by both the optimum detector and the radiometer is:
\begin{align}
p_{e,\text{det}}^{(w)}&=Q\left(\frac{\sqrt{n_p}\Delta\mu}{\sigma_0+\sigma_1}\right)+\frac{\epsilon_{n_p,0}+\epsilon_{n_p,1}}{2},\label{eq:pedet}
\end{align}
with the distinction arising from the expressions for the moments being different for the detectors.

\subsection{Discussion}
\label{app:detector_asymptotics}
When Alice follows the SRL, for both detectors, per Appendix \ref{ap:moments}, $\Delta\mu\propto\frac{1}{\sqrt{n_p}}$ and $\sigma_1\to\sigma_0$ as $n_p\to\infty$.
Thus, $p_{e,\text{det}}^{(w)}$ converges to a constant, meaning neither detector is effective.
On the other hand, when Alice does not follow the SRL, $p_{e,\text{det}}^{(w)}\to 0$ as $n_p\to\infty$, since $\sigma_0+\sigma_1$ are upper-bounded by a constant, while $\sqrt{n_p}\Delta\mu$ is unbounded and $\epsilon_{n_p,0},\epsilon_{n_p,1}\to 0$.
This is borne out in our experiments.

\revFinal{Furthermore, since $p_{e,\text{det}}^{(w)}$ converges to $p_e^{(w)}$ in \eqref{eq:tvbound} for optimal detector as $n_p\to\infty$, Taylor series expansion of \eqref{eq:pedet} around $\alpha_n=0$ yields \eqref{eq:anreqimproved}, which translates into $\approx 25\%$ improvement in $\alpha_n$ for $\delta=0.07$ used here.}

\section{Moments of Willie's Detectors' Sufficient Statistics}\label{ap:moments}
To analyze the error probability, we require the mean, variance, and upper bounds on the third absolute central moment of $S_i^{\text{(opt)}}$ and $S_i^{\text{(rad)}}$ under both hypotheses.

\subsection{Optimal Detector}
Under $H_0$, observations follow complex Gaussian distributions $\vec{w}_i=(\vec{p}_i, \vec{q}_i) \sim \mathcal{CN}(\vec 0, 2\sigma_w^2 \mathbf{I}_{n_s})$. Consequently, $S_i^{\text{(opt)}}$ is a sum of two independent exponential random variables. The mean $\mu_0$ and variance $\sigma_0^2$ are
\begin{align}
\mu_0 &= 2\sigma_w^2\left(\|\vec{c}_p\|^2 + \|\vec{c}_q\|^2\right), \label{eq:mu0}\\
\sigma_0^2 &= 4\sigma_w^4\left(\|\vec{c}_p\|^4 + \|\vec{c}_q\|^4\right). \label{eq:sigma0}
\end{align}
The Berry-Esseen bound relies on the third absolute central moment $\rho_0 = \mathbb{E}\left[|S_i - \mu_0|^3 \right]$.
We write $S_i - \mu_0 = (X - \mu_x) + (Y - \mu_y)$, where $X$ and $Y$ are independent exponential variables corresponding to the pilot and QPSK projections. Using the Minkowski inequality for the $L_3$ norm
\begin{align}
\rho_0 &= \| (X - \mu_x) + (Y - \mu_y) \|_3^3 \\&\leq \left( \| X - \mu_x \|_3 + \| Y - \mu_y \|_3 \right)^3.
\end{align}
For an exponential variable $X$, $\| X - \mu_x \|_3 = \mu_x (12/e - 2)^{1/3}$. Substituting this yields an upper bound for $\rho_0$
\begin{align}
\rho_0 \leq \left( \frac{12}{e} - 2 \right) (\mu_x + \mu_y)^3 = \left( \frac{12}{e} - 2 \right) \mu_0^3 \approx 2.415 \mu_0^3. \label{eq:rho0}
\end{align}

Under $H_1$, Willie receives a statistical mixture: with probability $1-\alpha_n$, the slot contains only noise ($t=0$), and with probability $\alpha_n$, it contains signal ($t=1$).
The conditional mean given $t=1$ is shifted by the signal energy $\Delta P = a_w^2\left(\|\vec{c}_p\|^4 + \|\vec{c}_q\|^4\right)$
\begin{align}
\mu_P = \mathbb E[S_i | t=1] &= \mu_0 + \Delta P.
\end{align}
The unconditional mean under $H_1$ is $\mu_1 = \mu_0 + \alpha_n(\mu_P - \mu_0) = \mu_0 + \alpha_n \Delta P$. Note that the difference in means is given by $\Delta \mu = \alpha_n \Delta P$.

The conditional variance given $t=1$ involves non-central terms
\begin{align}
\sigma_P^2 = \mathrm{Var}[S_i | t=1] &= \sigma_0^2 + 4\sigma_w^2 a_w^2 \left(\|\vec{c}_p\|^6 + \|\vec{c}_q\|^6\right).
\end{align}
Using the law of total variance, the unconditional variance $\sigma_1^2$ is
\begin{align}
\sigma_1^2 &= \sigma_0^2 + \alpha_n(\sigma_P^2 - \sigma_0^2) + \alpha_n(1-\alpha_n)(\Delta P)^2.
\end{align}

Analytically evaluating the third absolute central moment $\rho_1$ for the mixture distribution is intractable due to the absolute value function. Instead, we bound $\rho_1$ using the fourth central moment $\mu_{4,1}$ via Lyapunov's inequality $\rho_1 \leq \left( \mu_{4,1} \right)^{3/4}$.
The fourth central moment is derived algebraically from the raw moments. Let $m_k^{(0)} = \mathbb{E}\left[S_i^k \middle| t=0\right]$ and $m_k^{(1)} = \mathbb{E}\left[S_i^k \middle| t=1\right]$. The mixture raw moments are
\begin{align}
\mathbb{E}[S_i^k] = (1-\alpha_n)m_k^{(0)} + \alpha_n m_k^{(1)}.\label{eq:mom_raw}
\end{align}
From these, $\mu_{4,1}$ is obtained via the expansion \begin{align}
    \mu_{4,1} = \mathbb{E}[S_i^4] - 4\mu_1\mathbb{E}[S_i^3] + 6\mu_1^2\mathbb{E}[S_i^2] - 3\mu_1^4\label{eq:mu41}
\end{align}
and raw moments \cite[Eq.~following (29.31)]{JohnsonKotz1994} 
\begin{align}
m_k^{(1)} &= \sum_{j=0}^k \binom{k}{j} \left[ j! \beta_p^j L_j\left( -\frac{\lambda_p}{\beta_p} \right) \right]\\&\phantom{======}\times \left[ (k-j)! \beta_q^{k-j} L_{k-j}\left( -\frac{\lambda_q}{\beta_q} \right) \right],
\end{align}
where $L_k(u)$ are Laguerre polynomials and $\beta_p = 2\sigma_w^2\|\vec{c}_p\|^2$ and $\lambda_p = a_w^2\|\vec{c}_p\|^4$ denote the noise power and signal energy for the pilot (similarly $\beta_q= 2\sigma_w^2\|\vec{c}_q\|^2, \lambda_q=a_w^2\|\vec{c}_q\|^4$ for QPSK).

\subsection{Power Detector}
Under $H_0$, $\vec{w}_i\sim\mathcal{CN}(\vec{0},2\sigma^2_w\mathbf{I}_{n_s})$. Consequently,  $S_i^{\text{(rad)}}=\|\vec{w}_i\|^2$ is a scaled central $\chi^2$ random variable with $2n_s$ degrees of freedom and scaling factor $\sigma_w^2$. The mean and variance are 
\begin{align}
    \mu_0&=2 n_s \sigma_w^2, \\
    \sigma_0^2&=4 n_s \sigma_w^4.
\end{align}
We again bound the third absolute central moment $\rho_0$ via Lyapunov's inequality $\rho_0\leq(\mu_{4,0})^{3/4}$. The fourth central moment is given by 
\begin{align}
    \mu_{4,0}&= 48 n_s \sigma_w^8 (n_s + 2).
\end{align}
Under $H_1$, the detector statistic is drawn from a mixture distribution. Conditioned on signal presence ($t=1$), $S_i^{\text{(rad)}}$ follows a scaled non-central $\chi^2$ distribution with $2n_s$ degrees of freedom and non-centrality parameter $\lambda=\frac{\Delta P}{\sigma_w^2}$, with $\Delta P = a_w^2\left(\|\vec{c}_p\|^2+ \|\vec{c}_q\|^2\right)$. The conditional mean is then $\mu_P=\mu_0+\Delta P$, and the unconditional mean is given by 
\begin{align}
    \mu_1 &=\mu_0+\alpha_n\Delta P.
\end{align}
The conditional variance is $\sigma_P^2=\sigma_0^2 + 4 \sigma_w^2 \Delta P$. Applying the law of total variance, we obtain
\begin{align}
    \sigma_1^2&= \sigma_0^2+\alpha_n(\sigma_P^2-\sigma_0^2)+\alpha_n(1-\alpha_n)\Delta P^2 \\
    &=\sigma_0^2+4\alpha_n\sigma_w^2\Delta P + \alpha_n(1-\alpha_n)\Delta P^2.
\end{align}
We find the fourth central moment $\mu_{4,1}$ by expanding the moments of the mixture distribution. Let $\Delta\mu=\mu_1-\mu_0=\alpha_n \Delta P$ and $\Delta \mu'=\mu_P-\mu_1$. Now, we have 
\begin{align}
    \mu_{4,1} &= (1-\alpha_n) \mathbb{E}\left[\left(S_i^{\text{(rad)}}-\mu_0-\Delta\mu\right)^4\middle|t=0\right] \nonumber \\ &\phantom{=}+\alpha_n\mathbb{E}\left[\left(S_i^{\text{(rad)}}-\mu_P+\Delta\mu'\right)^4\middle|t=1\right].
\end{align}
Expanding the centered moments, the linear terms disappear, leaving 
\begin{align}
    \mu_{4,1} &= (1-\alpha_n) \left[ \mu_{4,0} - 4\Delta\mu \mu_{3,0} + 6(\Delta\mu)^2 \sigma_0^2 + (\Delta\mu)^4 \right] \nonumber \\
    &\phantom{=} + \alpha_n \left[ \mu_{4,P} + 4\Delta\mu' \mu_{3,P} + 6(\Delta\mu')^2 \sigma_P^2 + (\Delta\mu')^4 \right].
\end{align}
The third central moment for $t=0$ is $\mu_{3,0}=16 n_s\sigma_w^6$. For $t=1$, the moments are given by the cumulants of the scaled non-central $\chi^2$ distribution \cite[Ch.~29]{JohnsonKotz1994} as 
\begin{align}
    \mu_{3,P} &= 8\sigma_w^6 \left(2 n_s + 3\frac{\Delta P}{\sigma_w^2}\right)\\
    \mu_{4,P} &= 48\sigma_w^8 \left(2 n_s + 4\frac{\Delta P}{\sigma_w^2}\right) + 3 \sigma_P^4.
\end{align}

\section{Estimation of Willie's Channel Parameters}\label{ap:snr}
We need to estimate Willie's channel parameters from his observation and  knowledge of $\vec{c}_p$ and $\vec{c}_q$.
An empty pulse slot received by Willie contains only noise:
\begin{align}
\vec{w}_{p}(0)&=\vec{z}^{(w)},
\end{align}
The expectation of its squared magnitude is:
\begin{align}
    \mathbb{E}\left[\langle \vec{w}_{p}(0),\vec{w}_{p}(0)\rangle\right]&=\mathbb{E}\left[\langle \vec{z}^{(w)},\vec{z}^{(w)}\rangle\right]=2n_s\sigma_w^2.
\end{align}
Thus, averaging over many instances of empty pulse slots and dividing by a known constant $2n_s$ yields an estimate $\hat{\sigma}_w^2$ of $\sigma_w^2$.
This can be used to estimate $\mu_0$ and $\sigma_0^2$ for both the optimal detector and the radiometer, as per Appendix \ref{ap:moments}.
Similarly, one can estimate $a_w^2$ from the pulse slots used for transmission.

Now, denote the test statistic for either detector under hypothesis $H_h$, $h\in\{0,1\}$ by $S^{(h)}$. After normalizing, $\frac{S^{(h)}}{n_p}\to \mathbb{E}\left[S^{(h)}_i\right]$ by Law of Large Numbers. Arithmetic and estimate of $\sigma_w^2$ yield the quantities we need to calculate the threshold.
\bibliographystyle{IEEEtran}

\bibliography{papers, references}

\end{document}